\documentclass{aa}  
\usepackage{multirow}
\usepackage{graphicx}
\usepackage{txfonts}
\begin{document}

   \title{Influence of the optical Fe II quasi-continuum on measuring the spectral parameters of active galactic nuclei }

   \author{Luka \v C. Popovi\'c
   \inst{1,2},
         Jelena Kova\v cevi\' c-Doj\v cinovi\'c
         \inst{1},
         Ivan Doj\v cinovi\'c
         \inst{3}
          \and
          Ma\v sa Laki\' cevi\' c
          \inst{1}
          }

   \institute{Astronomical Observatory, Volgina 7, 11160 Belgrade, Serbia\\
              \email{lpopovic@aob.bg.ac.rs; jkovacevic@aob.bg.ac.rs}
         \and           
       Department of Astronomy, Faculty of Mathematics, University of Belgrade, Studentski Trg 16, 11000 Belgrade, Serbia\\
                        \and
    Faculty of Physics, University of Belgrade, Studentski Trg 12, 11000 Belgrade, Serbia\\
       }

%   \date{Received 19 February 2020 / Accepted 28 March 2020}

% \abstract{}{}{}{}{} 
% 5 {} token are mandatory

  \abstract
  {}
 {We explore the influence of optical Fe II quasi-continuum on the measured spectral parameters in the $\lambda\lambda$4150-5500 \AA \ range for the spectra of  Type 1 active galactic nuclei (AGNs).}
{We assume that the broad line region is composed of two sub-regions: the very broad line region (VBLR) and the intermediate line region (ILR). We constructed a large set of synthetic AGN spectra by taking different portions of the VBLR and ILR contributions, where initially the VBLR and ILR model spectra were constructed on the basis of prototypes of two observed spectra with dominant VBLR (i.e. ILR) emission. To investigate the influence of the optical Fe II quasi-continuum on the AGN measured spectral parameters, we fit the power-law continuum and emission lines in a set of model spectra, as commonly done for observed AGN spectra. We then compared the spectral parameters obtained after the fitting procedure with those of the model.}
{  We find that the optical Fe II quasi-continuum can be very strong in the case of spectra with strong and very broad Fe II lines and it is difficult to fully separate it from the power-law continuum. This gives the effect of a slightly underestimated H$\beta$ width and underestimated fluxes of the H$\beta$ and Fe II lines, while the continuum flux is then slightly overestimated. The most affected spectral parameters are the line equivalent widths (EWs), especially EW Fe II, which may be strongly underestimated. We discuss the possible underlying physics in the quasar main sequence, as implied by the results of our spectral modelling. We find that the set of AGN model spectra assuming different ILR and VBLR contributions can aptly reproduce the quasar main sequence, that is, the full width at half maximum (FWHM) H$\beta$ versus Fe II/H$\beta$ anti-correlation, where both parameters in this anti-correlation are strongly dependent on the  ILR and VBLR contribution rate. }
{}

    \titlerunning{ Influence of the optical Fe II quasi-continuum on measuring the spectral parameters of AGN}
               \authorrunning{Popovi\'c et al.}                                      

   \maketitle
%
%-------------------------------------------------------------------

\section{Introduction}\label{1}

Active galactic nuclei (AGNs) represent the most powerful sources in the universe, emitting energy across a broad wavelength band. They are divided into several groups that  exhibit a range of different spectral characteristics \citep[see e.g.][etc.]{pe97,of06,ne13}. Type 1 AGNs show different  broad lines in the UV/optical spectral range, which originate very close to the central supermassive black hole (SMBH) and can offer important information on the physics in the SMBH vicinity. Additionally, broad emission lines have commonly been used for making supermassive black hole (SMBH) mass estimates  \citep[for review, see][and reference therein]{pop20}. 

One of the important parts of the spectral region is the optical one around the H$\beta$ line. In particular, the H$\beta$ line and continuum at $\lambda$ 5100\AA\ are commonly used for single epoch mass estimates of SMBHs. Using the relation between mass and optical parameters, some other relations for SMBH mass estimations can be derived using the other lines apart from H$\beta$ (especially in the UV) and the corresponding continuum. Therefore, the spectral properties of the H$\beta$ and surrounding (continuum and line) spectra are very important for the SMBH mass determination, but also for the investigation of the other characteristics, such as AGN orientation, accretion rate, dust distribution around AGN, and so on \citep[see e.g.][etc.]{co06,jm06,la18,la22,sr22}.
 
 Additionally, in the H$\beta$ wavelength region, the emission of the Fe II lines is present \citep[see e.g.][etc.]{co80,jo81,kovacevic2010,sh10,ma16,pa22,ga22}, which is very important for the investigation of AGN physics. The origin of these lines and their characteristics have been investigated in a number of papers \citep[for review see][and reference therein]{ga22}. However, there are many open questions concerning the optical Fe II lines origin and their connection with other AGN properties \citep[see e.g.][]{kovacevic2015,le19,ma21}.  
 
 In principle, there are two distinct cases of AGN Type 1 spectra, one with generally narrower emission lines and stronger Fe II observed in the case of  Narrow Line Seyfert 1 (NLS1) galaxies, while and the other one shows the broad emission lines observed in the case of Broad Line Seyfert 1 (BLS1). The difference between these two types (with the exception of the optical Fe II strength), could be seen in the other  spectral properties \citep{su00}. \cite{Boroson1992} found that  as the optical to X-ray slope and equivalent width (EW) of the optical Fe II emission increase, the EW of the [OIII] lines (near H$\beta$)  decreases, and the full width at half maximum intensity (FWHM) of H$\beta$ also decreases. These  relationships represent the so-called eigenvector 1 (EV1)  correlations, obtained via a principal component analysis of  an AGN sample, which may result from a range of different effects. 
 
 It seems that the combination of the AGN orientation and accretion rate can cause different spectral properties of AGNs \citep{sh14}, indicating different physical AGN characteristics. The intensity of optical Fe II emission compared with the intensity of H$\beta$ can be a good indicator of physical processes in AGNs. In the plane of the FWHM H$\beta$ versus the intensity ratio of Fe II/H$\beta,$ there is  a trend showing that AGNs with a broader H$\beta$ have weaker  Fe II optical lines that make  so called  'quasar main sequence', which may be an indicator of the AGN accretion and orientation \citep[see][]{su00,sh14,ma18}. Using the optical Fe II strength relative to the H$\beta$ intensity and the FWHM H$\beta,$ it is possible to select different types of AGNs on the quasar main sequence. One of the divisions of these objects is based on population A and B AGNs (or quasars), taking into consideration that population A has FWHM H$\beta$ $<$ 4000 km s$^{-1}$ and strong Fe II emission \citep[see][and reference therein]{ma18}. Population B represents a group of AGNs with broader lines and weaker optical Fe II emission. However, there are some other spectral characteristics that can indicate different physical properties and optical Fe II origin \citep[as e.g. contribution of starbursts; see][]{pk11}.
 For example, \citet{du19} found a prominent correlation between the fluxes of Fe II and H$\beta$ emission lines with H$\beta$ lags, confirming an important role of accretion rate in driving the shortened lags between variability of the continuum at $\lambda$5100\AA\ and H$\beta$. Using this, they  established the scaling relation between the radius of the broad line region (BLR) and continuum luminosity using the relative strength of Fe II emission. This relation has been used for  estimates of the SMBH mass and accretion rate using the single-epoch spectra of AGNs.
 
 We may thus conclude that the optical Fe II lines are often used as an indicator of physical processes. Therefore, it is important to see different aspects of the Fe II origin and its influence on the spectral characteristics in the H$\beta$ wavelength region. One of the conclusions in \cite{kovacevic2010} was that the Fe II lines probably originate in the outer part of the BLR, but that a very broad Fe II component that originates closer to the SMBH may also be present. Due to the superposition of a number of the optical Fe II lines, it is possible that very broad Fe II components form a quasi-continuum, which is difficult to be distinguished from the real continuum emission. This motivates us to explore the influence of the optical Fe II quasi-continuum on AGN-measured spectral parameters, assuming that the BLR is complex and that Fe II and H$\beta$ emission can originate from two sub-regions of the BLR, one closer to the SMBH, emitting very broad lines, and one farther from the central SMBH, emitting narrower lines, which are typical for NLS1 AGNs. 
 
Therefore, for purpose of this research we constructed the set of synthetic AGN spectra and investigated the influence of the Fe II pseudo-continuum on the measurements of some spectral parameters.

 The paper is organised as follows. In Sect. \S2, we describe the theoretical basis of our model for BLR,  construction of set of synthetic, model spectra, and the method of extraction of the spectral properties. In Sect. \S3, we give our results and in Sect. \S4 we outline the conclusions.

\section{Method and theoretical base of modelling}\label{2}

In this section, we describe the two-component BLR model and the procedure for construction of synthetic spectra in 4150-5500 \AA \ spectral range, following this BLR model. Afterwards, we describe the fitting procedure of model spectra, in order to compare the spectral parameters obtained from the fit with those included in the model.

\begin{figure} 
 \centering
\includegraphics[width=85mm]{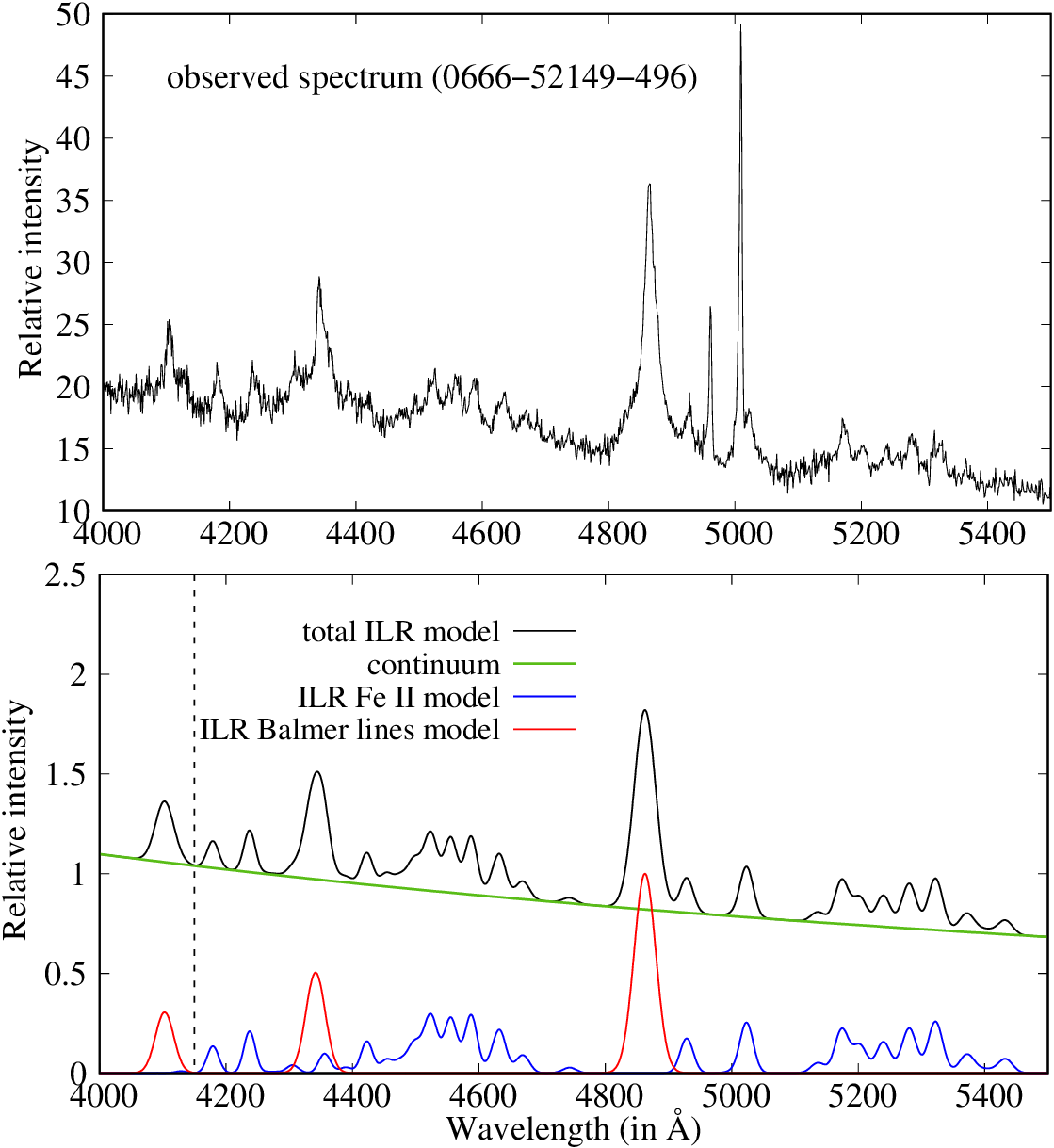}

\caption{Construction of the ILR model spectrum. Observed spectrum of SDSS J020039.16-084554.9 (SDSS plate-mjd-fiber: 0666-52149-496) after reddening and redshift correction {\it (top)}. ILR model spectrum made by using estimated parameters from the upper spectrum {\it (bottom)}. 
\label{fig1}}
 \end{figure}
 \begin{figure} 
 \centering
\includegraphics[width=85mm]{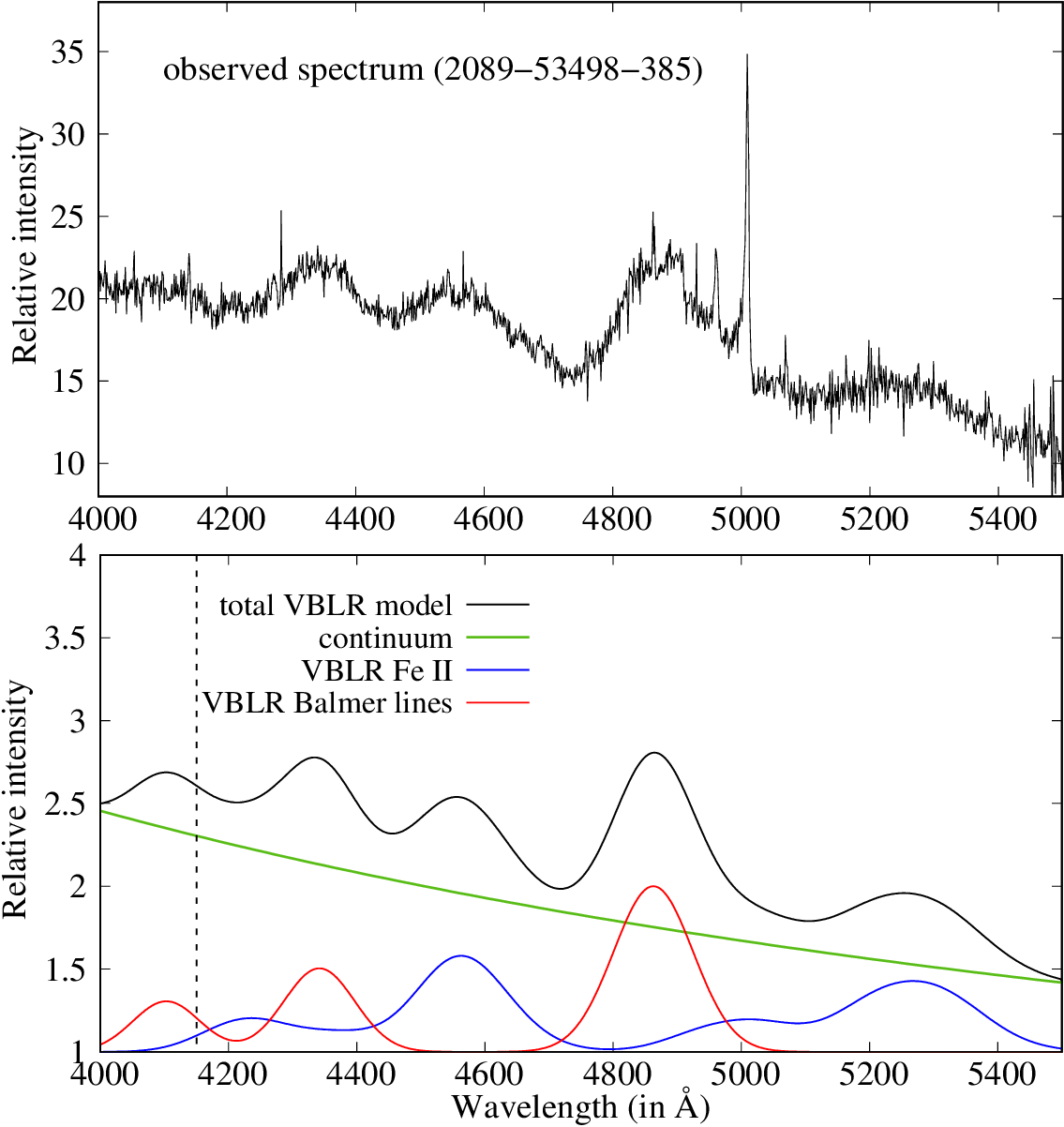}
\caption{Construction of the VBLR model spectrum. Details are the same as in Fig. \ref{fig1}, but for SDSS J120407.57+341916.3 (SDSS plate-mjd-fiber: 2089-53498-385) {\it (top)}, which is used for construction of the VBLR model spectrum {\it (bottom)}.
\label{fig2}}
 \end{figure}

\
\begin{figure*} 
 \centering
\includegraphics[width=150mm]{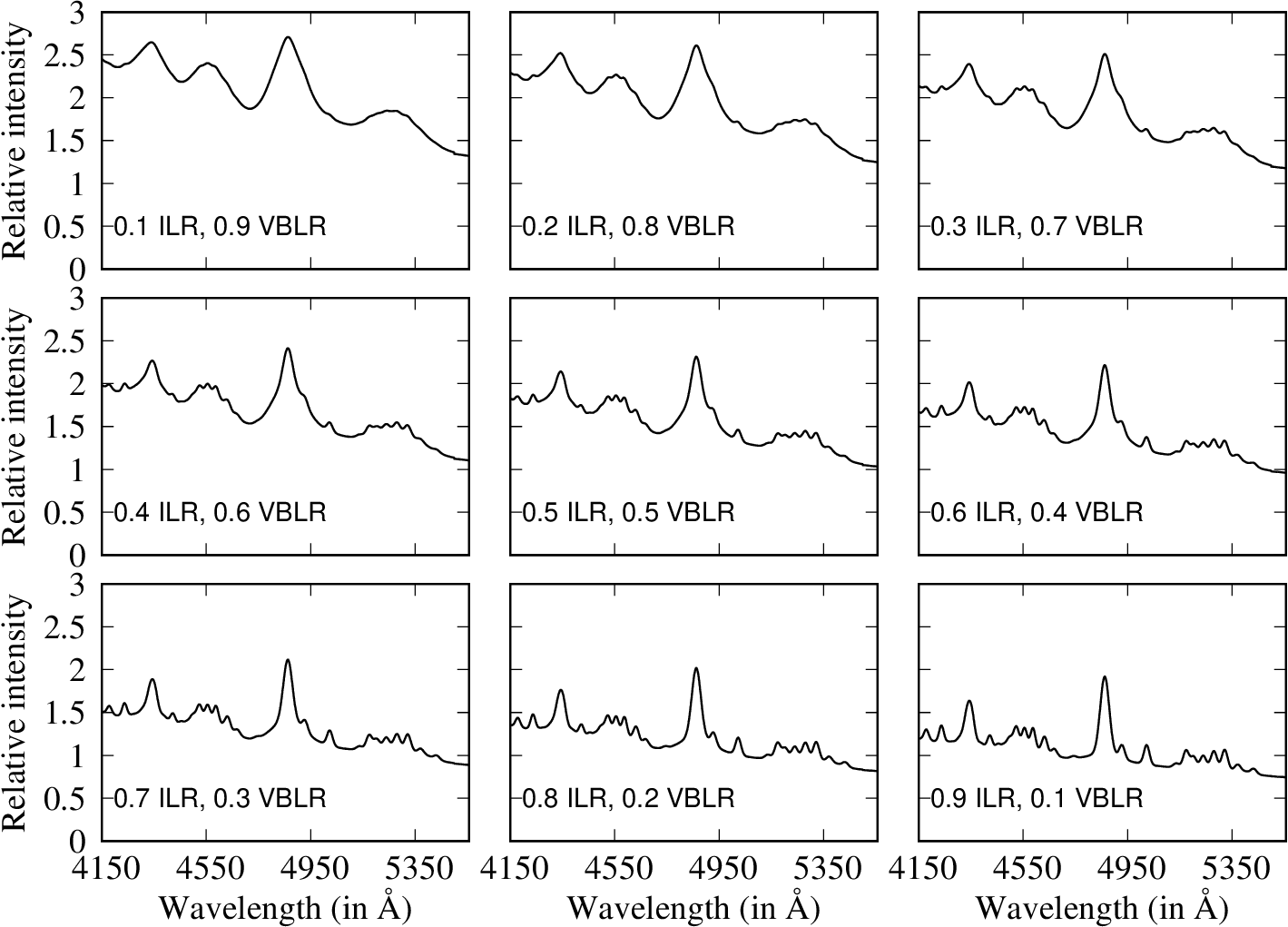}
\caption{Initial set of synthetic spectra made as linear combination of the ILR and VBLR model spectra.
\label{fig3}}
 \end{figure*}

\subsection{Two-component BLR model and optical Fe II emission}\label{2.1}

One of the open questions considering the Fe II emission is the place of origin with respect to the Fe II optical lines \citep[see e.g.][]{kovacevic2010,ba13,kovacevic2015}. As concluded in \citet{kovacevic2010}, the Fe II emission mostly originates from the outer part of the BLR,  but it seems that the very broad component, which originates from the inner part of the BLR, may be present too. The contribution of the very broad component of Fe II is difficult to estimate since it contributes to the quasi-continuum in the spectra. Therefore, here we consider the case where the optical Fe II lines originate in a complex emission region. We assume that the Fe II lines have a very broad component that is coming from the inner part of the broad line region, close to the central SMBH (i.e. the so-called very broad line region, VBLR), and a component that originates in an intermediate line region (ILR), which is an outer part of the broad line region. This model is the so-called two component BLR model, similar to the one accepted and discussed in a number of papers \citep[see e.g.][etc.]{po04,bo06,bo09,hu12,hu20}. This assumption is supported by several careful empirical analyses of Fe II lines, where it has been observed that Fe II lines probably have two components: one narrower and one broader \citep{VC2004,Dong2008,pa22}. As discussed in previous investigations based on a two-component BLR model, the contributions of the VBLR and ILR can be different in different spectra. To simplify the analysis, here we assume that the portion of VBLR contribution in Fe II lines is similar to that in Balmer lines in the same spectrum.
 
Analysing the properties of the Fe II lines in a large sample of Type 1 spectra where these lines have been carefully investigated  \citep[see the sample in][]{kovacevic2010, kovacevic2015}, we may notice two borderline cases: 1) the spectra with narrow Balmer lines and strong narrow Fe II lines (typical for NLS1s) and 2) the spectra with very broad Balmer lines and mostly weaker, broad Fe II lines. We assumed that the first ones are dominantly emitted from the ILR, while the second ones are likely to primarily be emitted from the VBLR. Therefore, first we constructed two basic model spectra, which are expected to represent the emission from ILR (i.e. VBLR) based on the properties of the two real prototype spectra, for which we assumed a dominant contribution from one of these two sub-regions.

 In order to build the models for the two mentioned borderline cases, we used the sample previously analysed by \citet{kovacevic2015}, where Fe II lines were fitted with the Fe II template described in \cite{kovacevic2010}. This template enables the estimation of the width of Fe II lines as well as the intensities of different Fe II line groups. Following the measurements of the line properties given in \cite{kovacevic2015}, we found two of the most extreme spectra, with the narrowest and broadest Fe II and H$\beta$ emission lines. Although the majority of the spectra with extremely broad Balmer lines have weak, broad Fe II lines, here we chose the broadest spectrum with a significant amount of Fe II emission to investigate the Fe II pseudo-continuum. The narrowest Fe II and H$\beta$ lines are seen in spectrum SDSS 0666-52149-496 (plate-mjd-fiber), where the FWHM of the broad H$\beta$ is equal to 1970 km s$^{-1}$ and FWHM of the Fe II lines is equal to 1500 km s$^{-1}$, as measured in \cite{kovacevic2015}. The relative intensity of H$\beta$ to Fe II 4549 \AA \ line is H$\beta$/Fe II 4549 \AA \ = 4.4.
The spectrum with the broadest Fe II and H$\beta$ from that sample is SDSS 2089-53498-385 (plate-mjd-fiber), where the FWHM of the broad H$\beta$ is equal to 9030 km s$^{-1}$ and FWHM of the Fe II lines is equal to 8950 km s$^{-1}$. The relative intensity of H$\beta$ to Fe II 4549 \AA \ line is H$\beta$/Fe II 4549 \AA \ = 1.8. These two spectra are shown in the upper panels of Figs. \ref{fig1}-\ref{fig2}.

We assumed that the emission from ILR (i.e. VBLR) is dominated in these two spectra, so we used them as prototypes for the construction of the ILR and VBLR model spectra.
The models are constructed by adding the flux of the continuum, the flux of optical Fe II lines, and the broad components of the Balmer lines (H$\beta$, H$\gamma$ and H$\delta$). The Fe II lines are reproduced using the Fe II template given in \cite{kovacevic2010}, extended for Fe II lines given in \cite{sh12}, where Fe II widths and relative intensities between Fe II line groups are taken to be as measured in prototype spectra. Balmer lines are represented with a single Gaussian for each line, in order to simplify the model. The H$\beta$ lines have the same FWHM and relative intensity to Fe II as measured in prototype spectra, while H$\gamma$ and H$\delta$ lines are taken to have the same width as H$\beta$ line, and their intensities relative to H$\beta$ are taken to be H$\beta$/H$\gamma$ = 0.5, and H$\delta$/H$\gamma$ = 0.3, following Case B \citep{of06}. We put that all considered lines in models have no shift relative to the referent wavelength. To simplify the analysis, the narrow lines (narrow Balmer lines and [O III] lines) and broad He II $\lambda$4687.01 \AA \ and He I $\lambda$4027.32 \AA \ lines are not included in the models. The continuum emission included in ILR and VBLR model spectra is the same as estimated power-law continuum level in prototype ILR and VBLR spectra. For the prototype ILR spectrum (SDSS 0666-52149-496) estimated continuum is: I${_{cont ILR}}(\lambda)$ = 10.9$\cdot$($\lambda$/5693.7)$^{-1.5}$, while in the prototype VBLR spectrum (SDSS 2089-53498-385), it is: I${_{cont VBLR}}(\lambda)$ = 9.8$\cdot$($\lambda$/5697.4)$^{-1.7}$.

Finally, the intensities of all considered lines and continuum levels are normalised to H$\beta$ intensity, so in both models the H$\beta$ intensity is equal to 1. In the model construction, we used the intensities measured in the two prototype spectra for the relative intensities of the lines included in the model, instead of arbitrary relative intensities, to achieve as many realistic models, since these ratios probably vary for spectra with different line widths. The models for the ILR and VBLR spectra are shown in the lower panels of Figs. \ref{fig1}-\ref{fig2}.

\begin{figure} 
 \centering
\includegraphics[width=87mm]{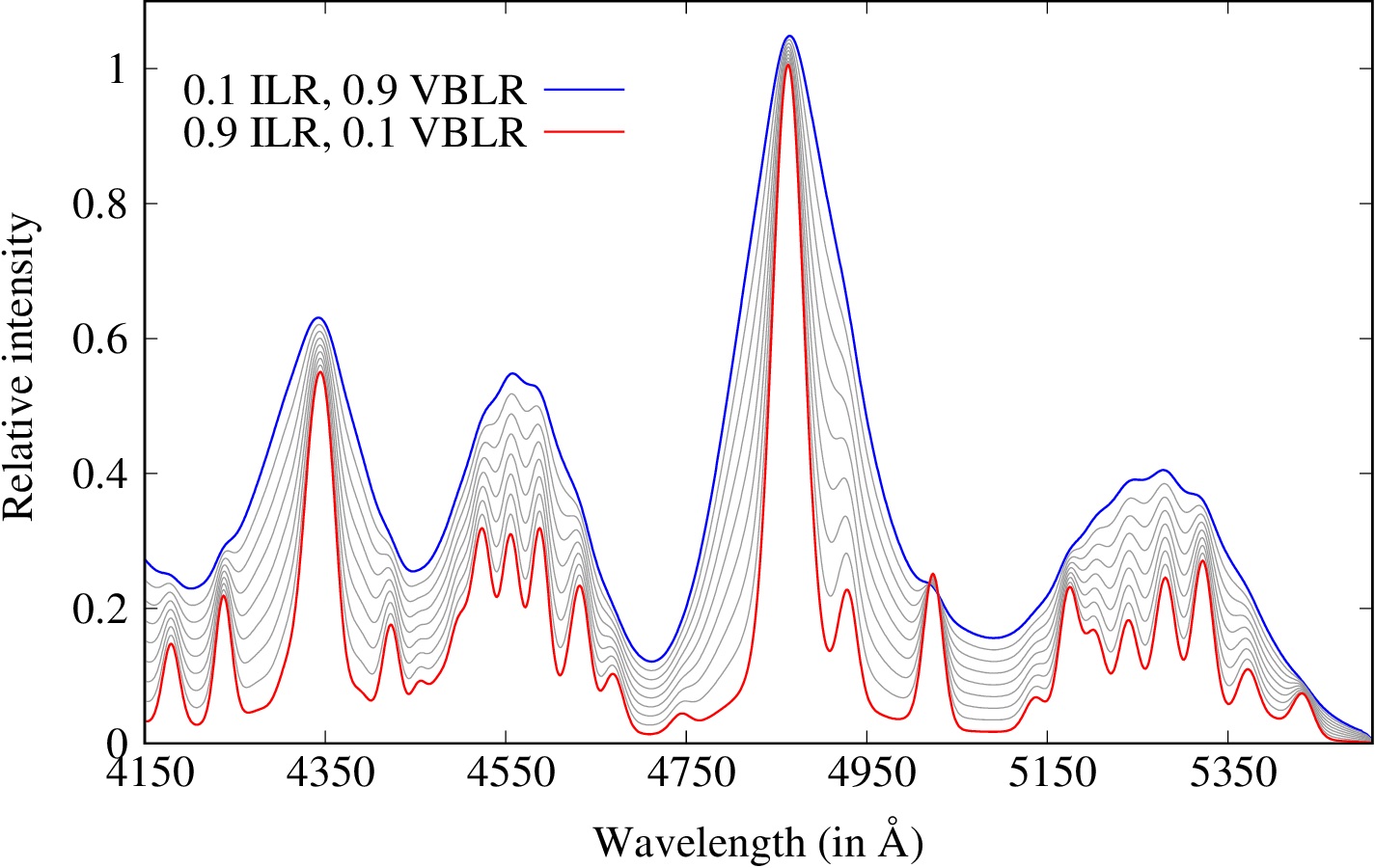}
\caption{Initial set of models (nine modelled spectra): the comparison of the emission lines (Balmer lines + Fe II), with no continuum included. The broadest model is VBLR:ILR equal to 0.9:0.1 (blue) and the narrowest is VBLR:ILR equal to 0.1:0.9 (red).
\label{fig4}}
 \end{figure}

 After obtaining the two initial ILR and VBLR spectral models, we created the initial set of synthetic spectra, obtained as a linear combination of the ILR and VBLR spectral models. In this way, we reproduced the multiple spectra with different contributions of the ILR and VBLR components in the Fe II and Balmer lines. Composite model spectra ($I_{comp}(\lambda)$) are defined as: 
 
 $$I_{comp}(\lambda)=p_1\cdot I_{ILR}(\lambda)+p_2\cdot I_{VBLR}(\lambda),$$
where $ I_{ILR}(\lambda)$ and $I_{VBLR}(\lambda)$ are spectra corresponding to the ILR and VBLR spectral models, respectively (shown in the lower panels of Figs. \ref{fig1}-\ref{fig2}) and $p_1, p_2$ are the rate of contribution of the ILR and VBLR, respectively, assuming that $p_1+p_2=1$. 

Since the ILR and VBLR model spectra are normalised to H$\beta$ maximum (the H$\beta$ intensity is equal to 1), when multiplying these two spectra with different coefficients, $p_i$, we obtain a summary spectrum where the maximum of H$\beta$ is equal to 1. In this way, we can explore line shapes and their ratio relative to the H$\beta$ intensity. The pairs of coefficients ($p_1, p_2$) take values from 0.1 to 0.9, with step 0.1 (the ratio of $p_1$:$p_2$ takes following values: 0.1:0.9, 0.2:0.8, 0.3:0.7,..., 0.8:0.2, 0.9:0.1). In this way, we constructed a set of nine model spectra with different contributions from the ILR and VBLR components.

The He I may significantly contribute to the pseudo-continuum level at $\sim$4000 \AA, therefore, in the further analysis, we use the spectra models in the range of 4150-5500 \AA \ (see vertical dashed line in Figs. \ref{fig1}-\ref{fig2}). In that range, there is no contribution from the broad He I line. However, in the case of spectra with very broad lines, the broad H$\delta$ contributes to the pseudo-continuum level at $\sim$4150 \AA \ (see Fig. \ref{fig2}), and therefore the flux of H$\delta$ is included in all models for wavelengths larger than $\sim$4150 \AA. The initial set of model spectra (continuum and emission lines included), obtained as a linear combination of the ILR and VBLR spectral models, are shown in Fig. \ref{fig3}. In Fig. \ref{fig4}, we compared only the emission lines (Balmer + Fe II lines) from the initial set of model spectra, without including continuum emission. It can be seen that in the case of the broadest model (ILR contribution 0.1, VBLR contribution 0.9),
the Fe II and Balmer lines are blended, creating the significant pseudo-continuum.

\begin{figure*} 
 \centering
\includegraphics[width=170mm]{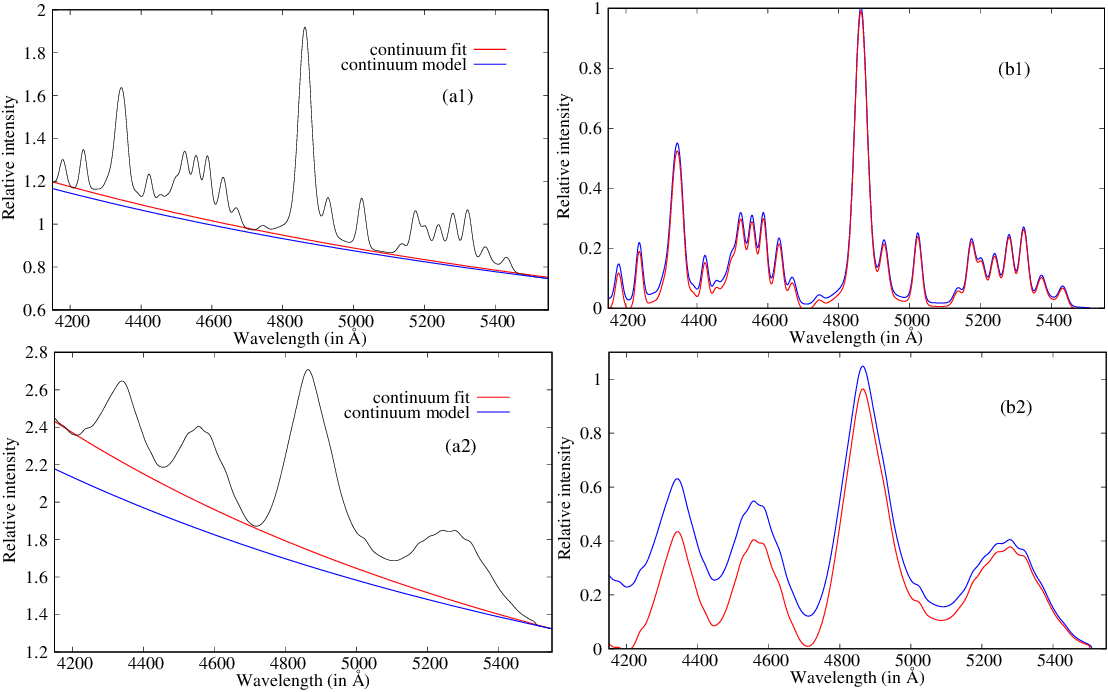}
\caption{ Comparison between the estimated continuum level by the fitting procedure and the initial continuum included in models. The initial model with 0.9 ILR and 0.1 VBLR contribution is given in panels (a1) and (b1), and the initial model with 0.1 ILR and 0.9 VBLR contribution is given in panels (a2) and (b2). In panels (a1) and (a2), the continuum level estimated by fitting is assigned the red line, while the continuum included in the model is assigned the blue line. In panels (b1) and (b2), the red line represents the line flux after subtraction of the continuum determined from fit, while the blue line is the sum of the H$\beta$+Fe II from model.
\label{fig5}}
 \end{figure*}

\subsection{Construction of the large set of model spectra}\label{2.2}

 Additionally, we created a large set of model spectra, giving more possibilities for width variation. In order to do that, we created several ILR and VBLR model spectra, with different widths of lines, which should be combined in order to create the larger sample of synthetic spectra. 
 
The width ranges for which we created ILR and VBLR models are determined following empirical results given in \cite{kovacevic2010}, performed by fitting a low-redshifted (z $<$ 0.7) sample of $\sim$300 AGNs Type 1, obtained from SDSS. In that work, the H$\beta$ line is fitted with two-component model, assuming that the core component arises in ILR and the wing component in VBLR. The measured Doppler widths of H$\beta$ ILR and VBLR components, obtained from the best fit, are given in \cite{kovacevic2010} in the Table 12. Following the ranges obtained for ILR and VBLR widths from this research, we obtained the six ILR model spectra and the 11 VBLR model spectra. The six ILR model spectra have the same continuum level and relative intensities between Fe II and Balmer lines as measured in SDSS 0666-52149-496, just for different FWHMs of all emission lines (Balmer lines and Fe II): 1500 km s$^{-1}$, 2000 km s$^{-1}$, 2500 km s$^{-1}$, 3000 km s$^{-1}$, 3500 km s$^{-1}$, and 4000 km s$^{-1}$. Similarly, we obtained the 11 VBLR model spectra with the same continuum level and relative intensities between Fe II and Balmer lines as measured in the initial VBLR model (SDSS 2089-53498-385), but including different FWHMs of all emission lines: 4500 km s$^{-1}$, 5000 km s$^{-1}$, 5500 km s$^{-1}$, 6000 km s$^{-1}$, 6500 km s$^{-1}$, 7000 km s$^{-1}$, 7500 km s$^{-1}$, 8000 km s$^{-1}$, 8500 km s$^{-1}$, 9000 km s$^{-1}$, and 9500 km s$^{-1}$. In this set of ILR and VBLR models, we assume that the ILR components have the same widths for Balmer lines and Fe II. The same is assumed for the VBLR components of these lines. This assumption is supported by the results given in \cite{kovacevic2010}, where significant correlations were found on a large sample between the width of Fe II lines and the widths of H$\beta$ ILR and H$\beta$ VBLR components.

The large set of model spectra is constructed similarly to the initial set of models, as described above. We performed the linear combination of each pair of ILR and VBLR models, with coefficients p1 and p2, where the coefficients may have values in the range of [0.1,0.9], with step 0.1, and p1$+$p2=1. Finally, this gives 594 synthetic spectra, with FWHM of total H$\beta$ (ILR+VBLR) in the range of 1500 - 8700 km s$^{-1}$. Since we assume that the portion of VBLR over ILR component is the same for H$\beta$ and Fe II lines and that the ILR and VBLR components of both lines have the same widths, the FWHM of Fe II lines is the same as for H$\beta$ in each spectrum and in the range of 1500 - 8700 km s$^{-1}$ for a large set of model spectra.

In further text, the nine synthetic spectra constructed as a linear combination of the ILR and VBLR models obtained directly from  the prototype, observed spectra will be called the "initial set of model spectra" (shown in Figs. \ref{fig3} and \ref{fig4}), while the set of 594 synthetic spectra created by variation in widths of the ILR and VBLR components will be called the "large set of model spectra."

\subsection{Fitting the synthetic spectra}\label{2.3}

In order to investigate the influence of the underlying Fe II and Balmer line pseudo-continuum on the measured parameters in AGN spectra, we performed the following test. We calculated several spectral parameters directly from the model: flux of the continuum at 5100 \AA, EWs and fluxes of H$\beta$ and Fe II lines, and FWHM H$\beta$. Afterwards, we determined the continuum level in these modelled spectra, as commonly done, by fitting power law to some continuum windows given in literature, as continuum windows at 4210-4230 \AA,  5080-5100 \AA, and 5600-5630 \AA \  \citep[see][]{Kuraszkiewicz2002}.

The difference between the estimated continuum level using continuum windows and the real continuum level included in the model is shown in Fig. \ref{fig5}. As it can be seen in Fig. \ref{fig5}, the difference between measured continuum and the modelled one is not large in the case of synthetic spectra with narrow lines, but in the case of spectra with very broad lines, where the contribution of VBLR dominates over the ILR, it can be significant.

After we obtained the best fit of the continuum, it was subtracted from the set of model spectra, and the Fe II and Balmer lines are fitted with the same procedure as described in \cite{kovacevic2015} for the optical range. The Fe II lines were fitted with the Fe II template given in \cite{kovacevic2010} and \cite{sh12}, and were described with nine free parameters:\ width, shift, intensities for six line groups, and the temperature required for the calculation of the relative intensities of the Fe II lines within one group. The Balmer lines were fitted with a double-Gaussian model: one Gaussian fits the core and the other fits the broad wings of the line. The intensities of all components in H$\beta$, H$\gamma,$ and H$\delta$, are the free parameters, while the widths and shifts of the core components of H$\beta$, H$\gamma,$ and H$\delta$ are the same, as are the widths and shifts of the wing components in these three lines.

As a result of the fit, we obtained the new values for the flux of continuum at 5100 \AA, EWs and fluxes of H$\beta$ and Fe II lines, and FWHM H$\beta$. This procedure was done together for the initial set of 9 model spectra and for the large set of 594 model spectra.

 \section{Results}\label{3}

First, we analysed how our set of model spectra is representative in parameter space of the large sample of the observed spectra. Then, we analysed the difference between spectral parameters obtained directly from the model and those measured after continuum subtraction and the fitting procedure.  Afterwards, we investigated the influence of the Fe II quasi-continuum, which is mostly coming from the VBLR, on the measured spectral properties that have been used in the investigation of different correlations. In most plots, we assigned in different ways the large set of 594 model spectra and the initial set of 9 model spectra.

\subsection{Quasar main sequence and set of model spectra}\label{3.1} 

\begin{figure} 
 \centering
 \includegraphics[width=0.48\textwidth]{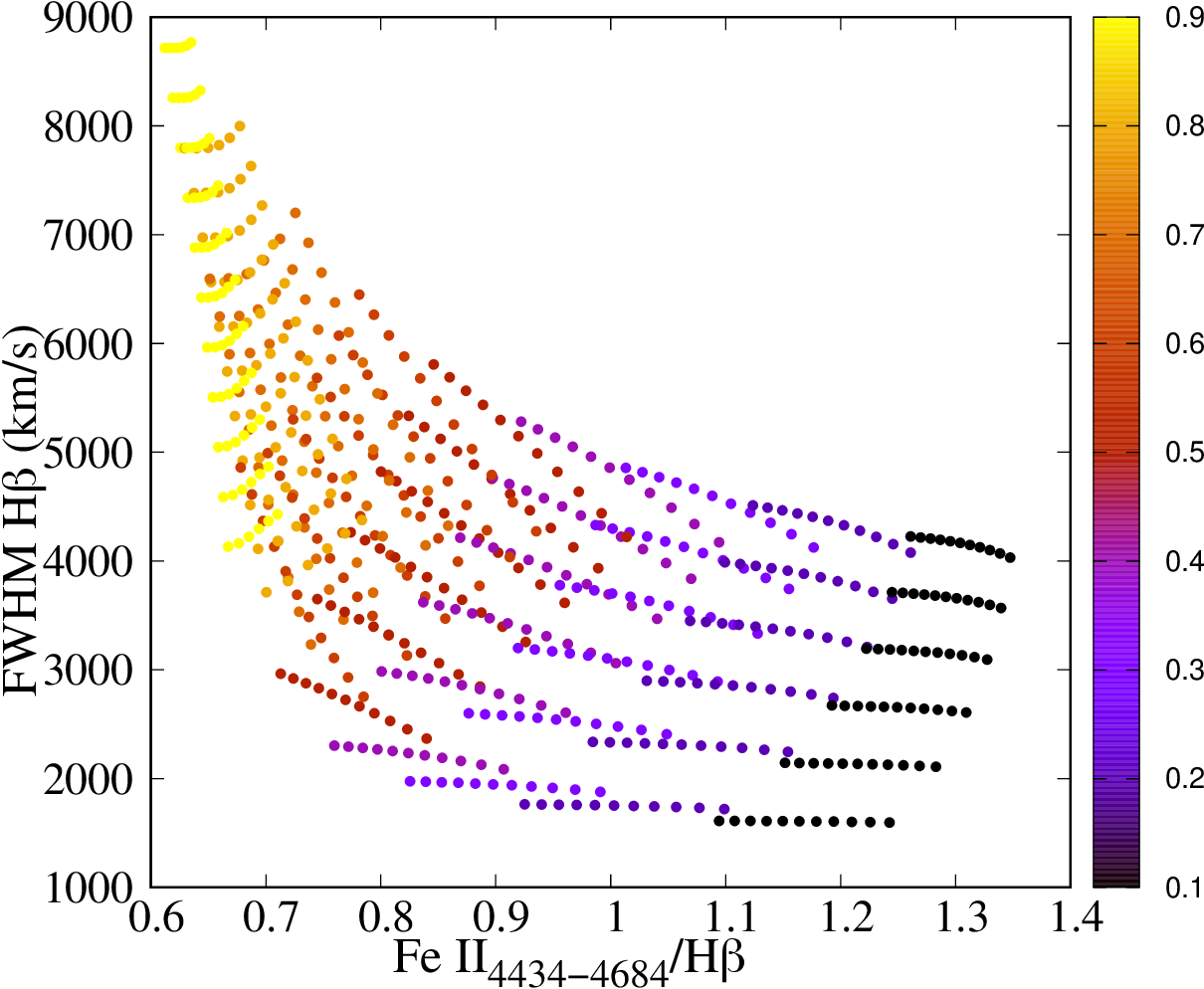}
\caption{ Correlation between FWHM H$\beta$ vs. Fe II/H$\beta$ (EV1 correlation) for a large set of 594 model spectra. The colour palette represents the contribution of the VBLR included in the model.
\label{fig5_1}}
 \end{figure}

\begin{figure} 
 \centering
 \includegraphics[width=0.48\textwidth]{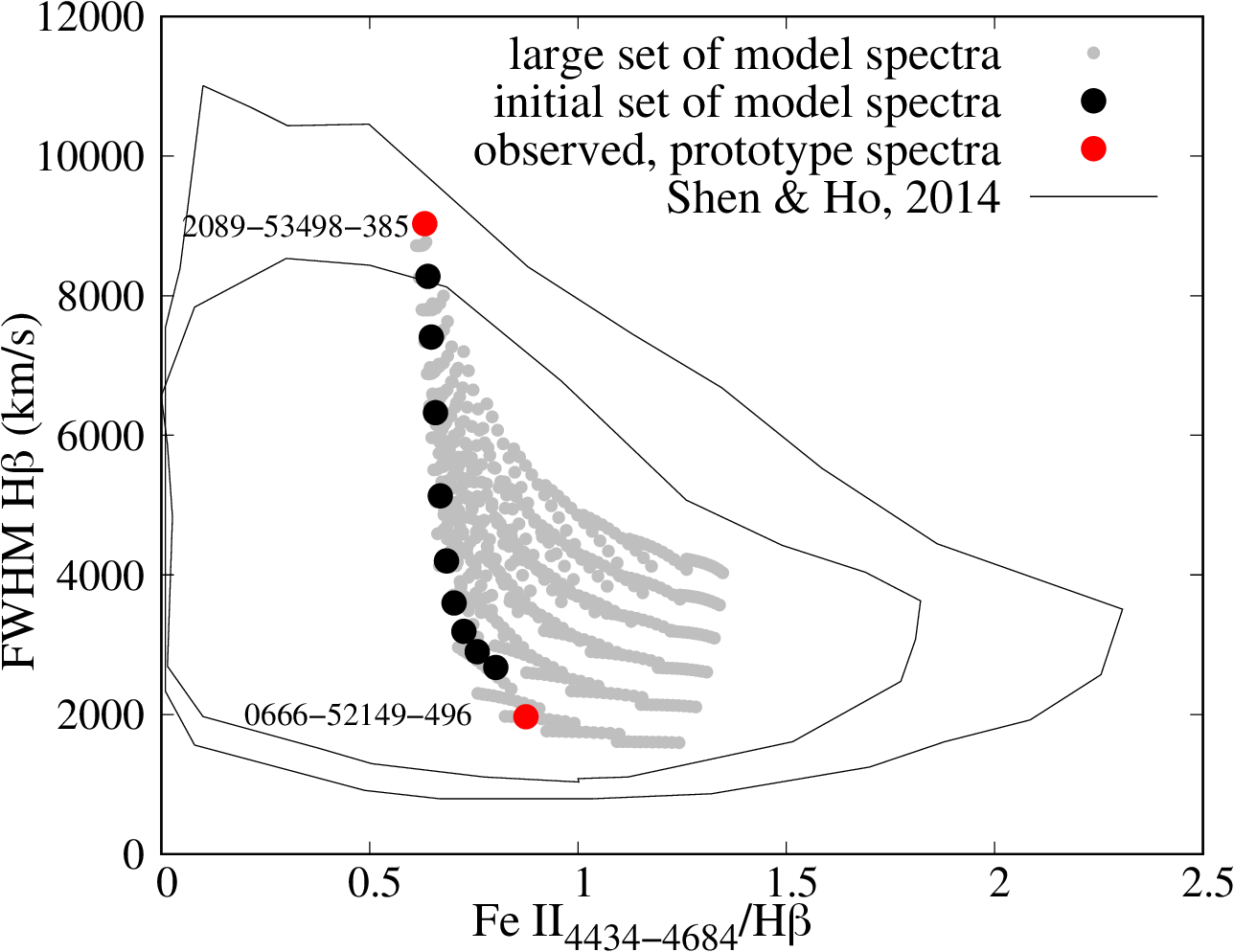}
\caption{ EV1 parameter space for the large set of 594 model spectra (grey dots) and the initial set of model spectra (black dots) compared with the contour of the total SDSS quasar sample from  \cite{sh14}. Two observed spectra used as prototypes for initial ILR and VBLR models (SDSS plate-mjd-fiber: 0666-52149-496 and 2089-53498-385, respectively) are designated with red dots.
\label{fig5_11}}
 \end{figure}

 The quasar main sequence is represented by the distribution of objects in the plane of the FWHM H$\beta$ versus FeII/H$\beta$ flux ratio, which, in principle, shows Eigenvector 1 correlations \citep{Boroson1992}. This is an important tool in understanding AGN properties \citep[see][]{sh14}. 
 To check whether our synthetic sample of spectra follow the quasar main sequence, we plotted the FWHM of H$\beta$ versus the ratio of Fe II flux (in the 4434-4684 \AA \ range) and flux H$\beta$, using parameters, as included in models. It is interesting to note that FWHM H$\beta$ versus Fe II/H$\beta$ anticorrelation is well reproduced with constructed set of spectral models, with different rates of ILR/VBLR contribution (see Fig. \ref{fig5_1}). In Fig. \ref{fig5_1}, the different colours of the dots represent different VBLR contributions in the model spectra (from 0.1 to 0.9). It can be seen that FeII/H$\beta$ flux ratio is strongly anticorrelated to the VBLR contribution (see colour palette), while the FWHM H$\beta$ correlates with it. In our set of model spectra, the majority of objects with FWHM H$\beta$ > 4000 km s$^{-1}$ (Pop B) have a VBLR contribution in the range of 0.5-0.9, while objects with FWHM H$\beta$ < 4000 km s$^{-1}$ (Pop A), have a VBLR contribution in the range of 0.1-0.5.

 The Fe II/H$\beta$ ratio of our set of models is in the range of 0.6-1.4, while the FWHM H$\beta$ is in the in range of 1500-9000 km s$^{-1}$. These values are within the range of Fe II/H$\beta$ and FWHM H$\beta$ parameters measured in large sample of about 20000 QSO spectra from SDSS in \cite{sh14} (see contour in Fig. \ref{fig5_11}). The FWHM H$\beta$ range of large QSO spectra in \cite{sh14} is 1500-11000 km s$^{-1}$, while the range of ratio Fe II/H$\beta$ is 0-2.3. The range of Fe II/H$\beta$ in our sample is narrower than one in the large QSO sample from \cite{sh14} since we adopted relative intensities of H$\beta$ and Fe II from only two prototype spectra (from SDSS 0666-52149-496 for all ILR spectral models, and from SDSS 2089-53498-385 for all VBLR spectral models). The linear combinations of ILR and VBLR spectral models result in different relative intensities for these lines in different model spectra, but they still do not cover all possibilities for the Fe II/H$\beta$ ratio which could be seen in real AGN spectra. Also, both our prototype spectra have strong Fe II lines relative to H$\beta$, so our models do not include the spectra with weak Fe II emission (as can be seen in Fig. \ref{fig5_11}). 
 
 \begin{figure} 
 \centering

\includegraphics[width=0.32\textwidth]{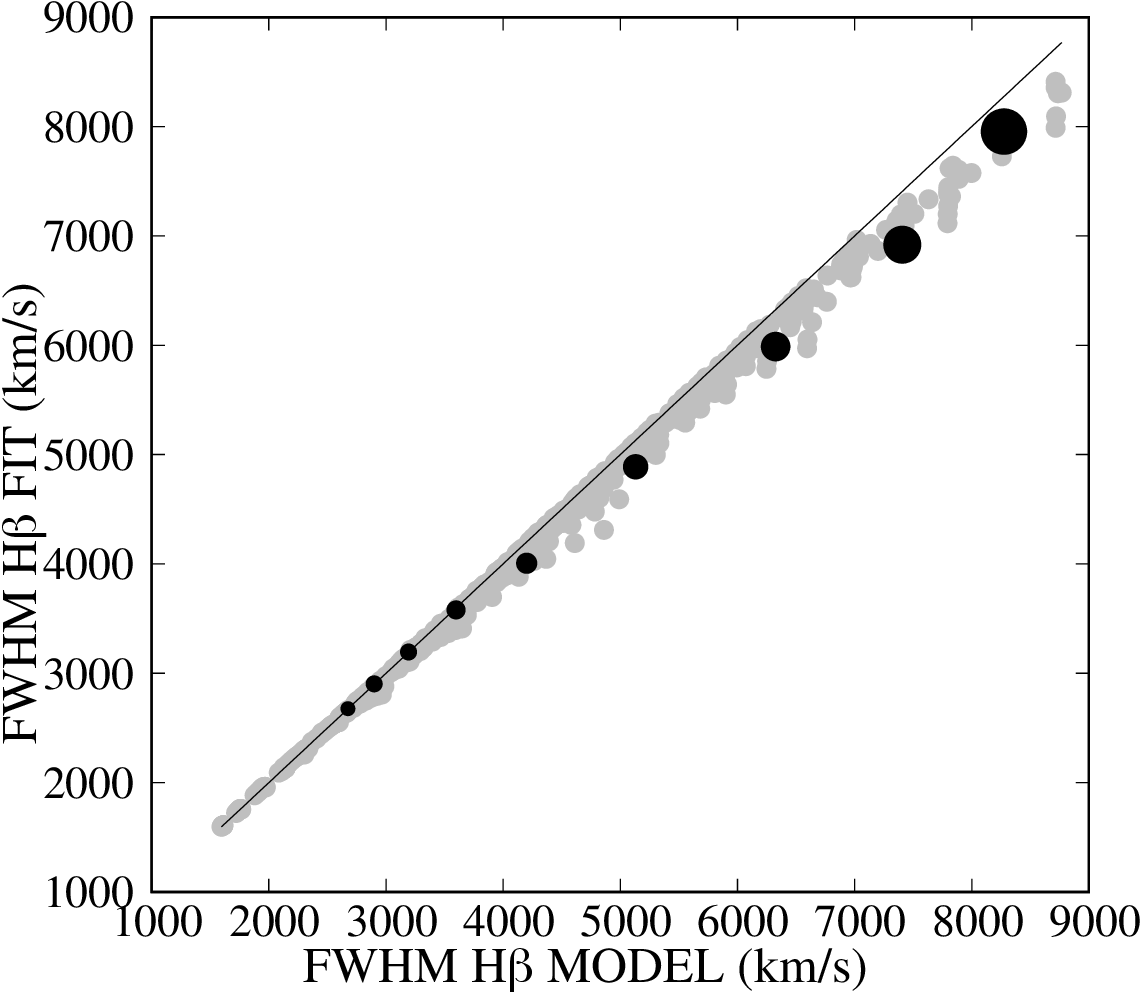}
\includegraphics[width=0.32\textwidth]{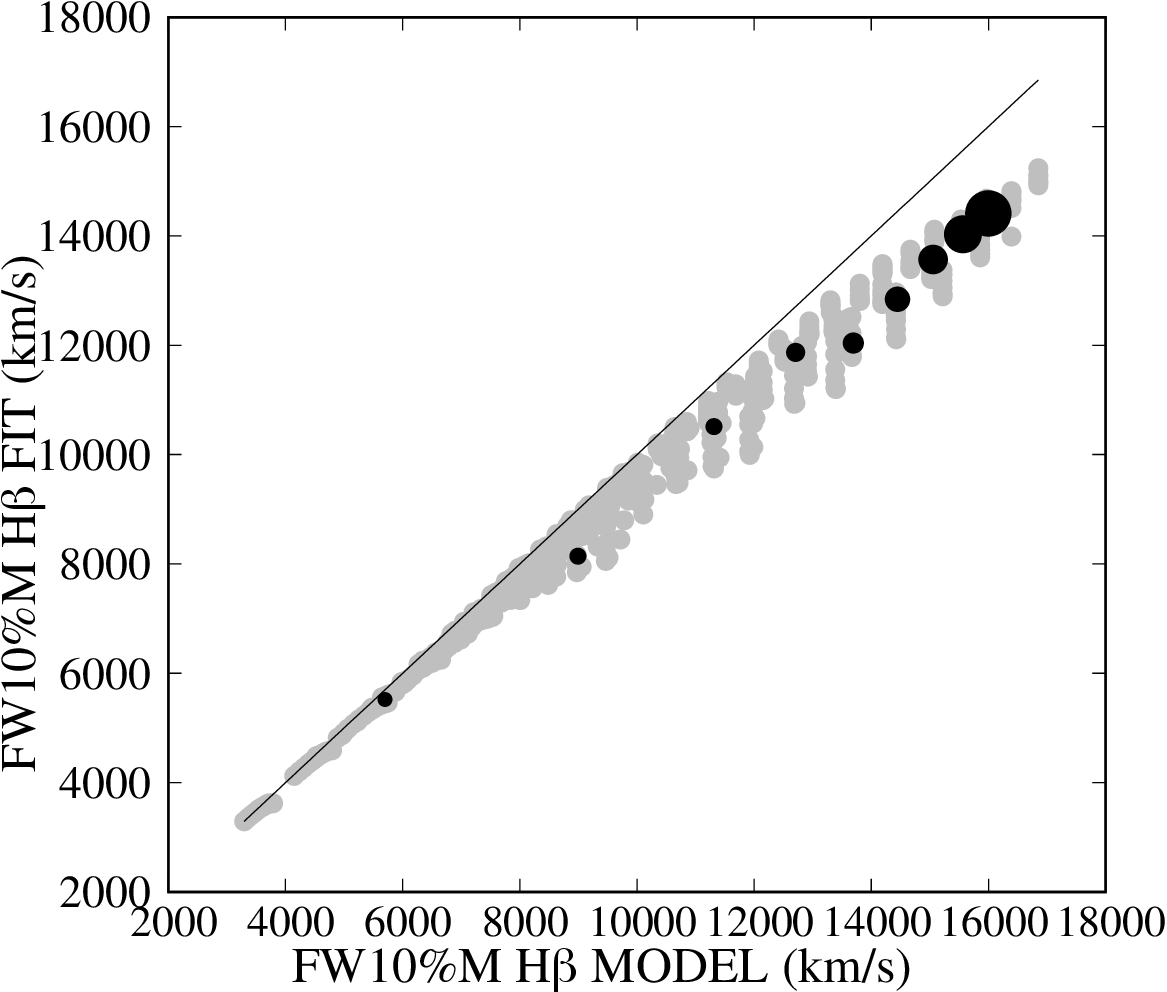}
\caption{Comparison between spectral parameters obtained from the best fit after continuum subtraction using continuum windows, with the same parameters as they are included in the model, for FWHM H$\beta$ {\it (top)} and FW10\%M H$\beta$ {\it (bottom)}. The black dots are values obtained from the initial set of model spectra, and their size increases as the VBLR contribution increases in the model spectra.
The grey dots are values obtained from the large sample of model spectra. The solid line displays a one-to-one relationship.
\label{fig6_0}}
 \end{figure}

\begin{figure} 
 \centering
\includegraphics[width=0.32\textwidth]{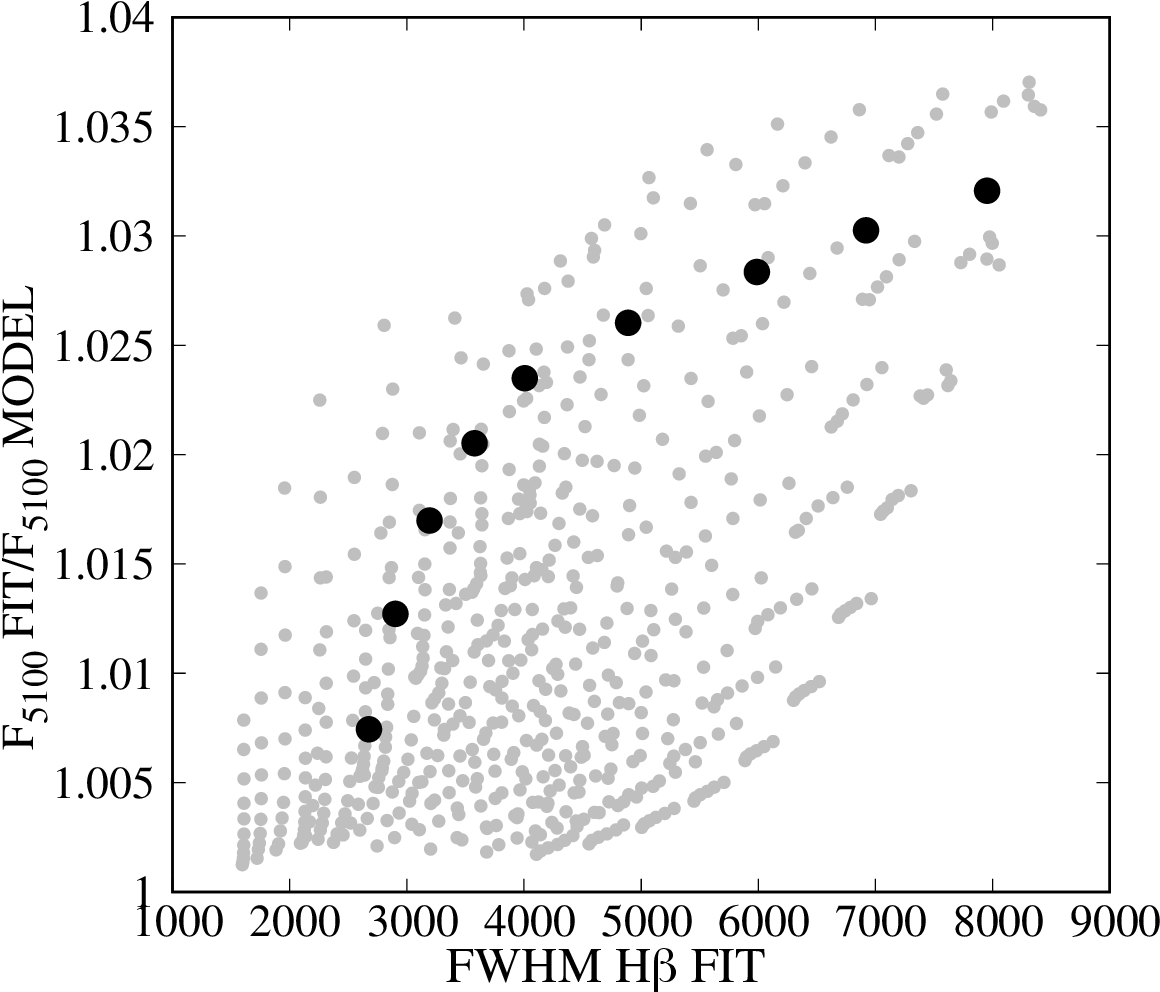}
\includegraphics[width=0.32\textwidth]{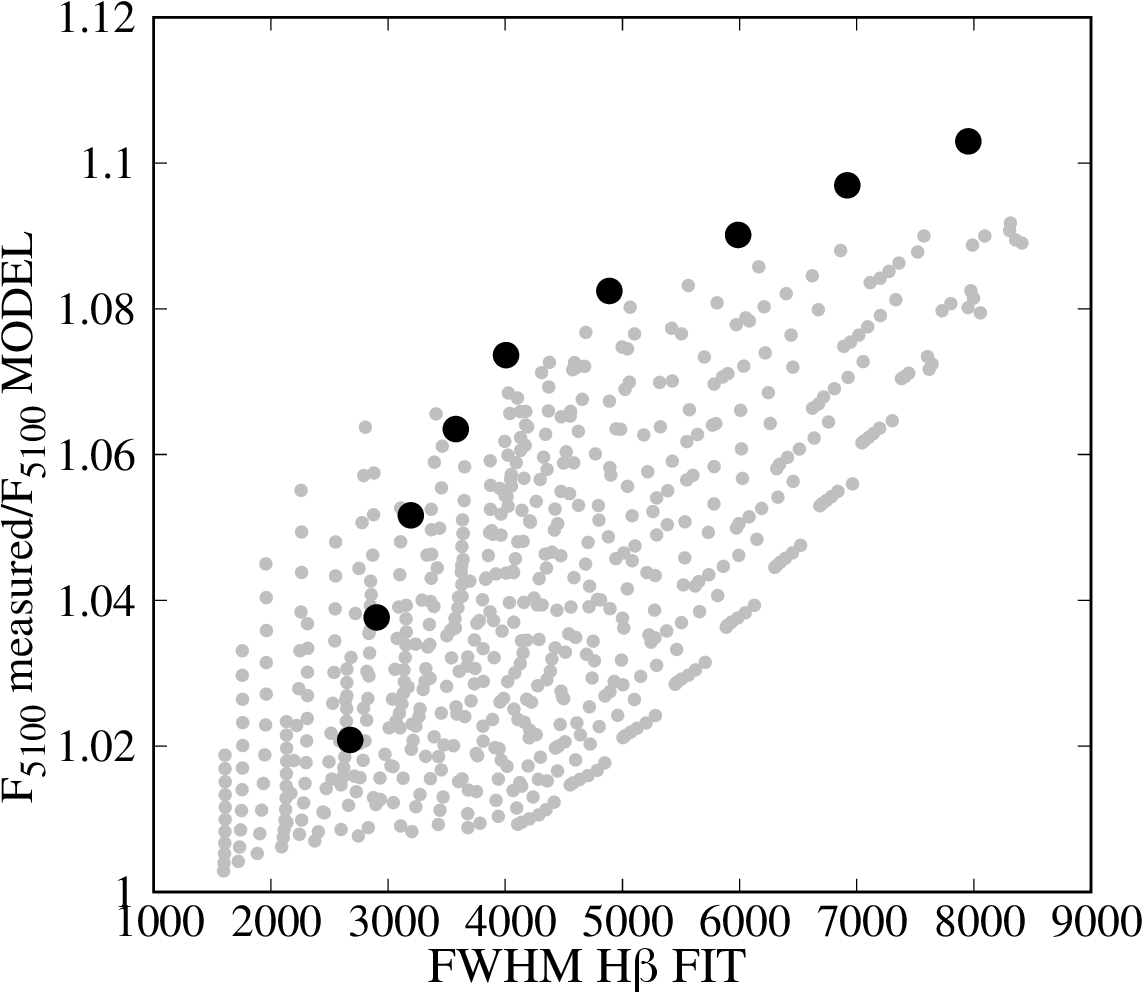}
\caption{Comparison between continuum flux at 5100 \AA \  from the model and obtained from the best fit of the power law {\it (top)} along with same, but for the continuum flux measured at  5100 \AA \ {\it (bottom)}. The black dots are values obtained from the initial set of model spectra, while the grey dots are values obtained from the large sample of model spectra.
\label{fig6_00}}
 \end{figure} 

\begin{figure*} 
 \centering
 \includegraphics[width=0.32\textwidth]{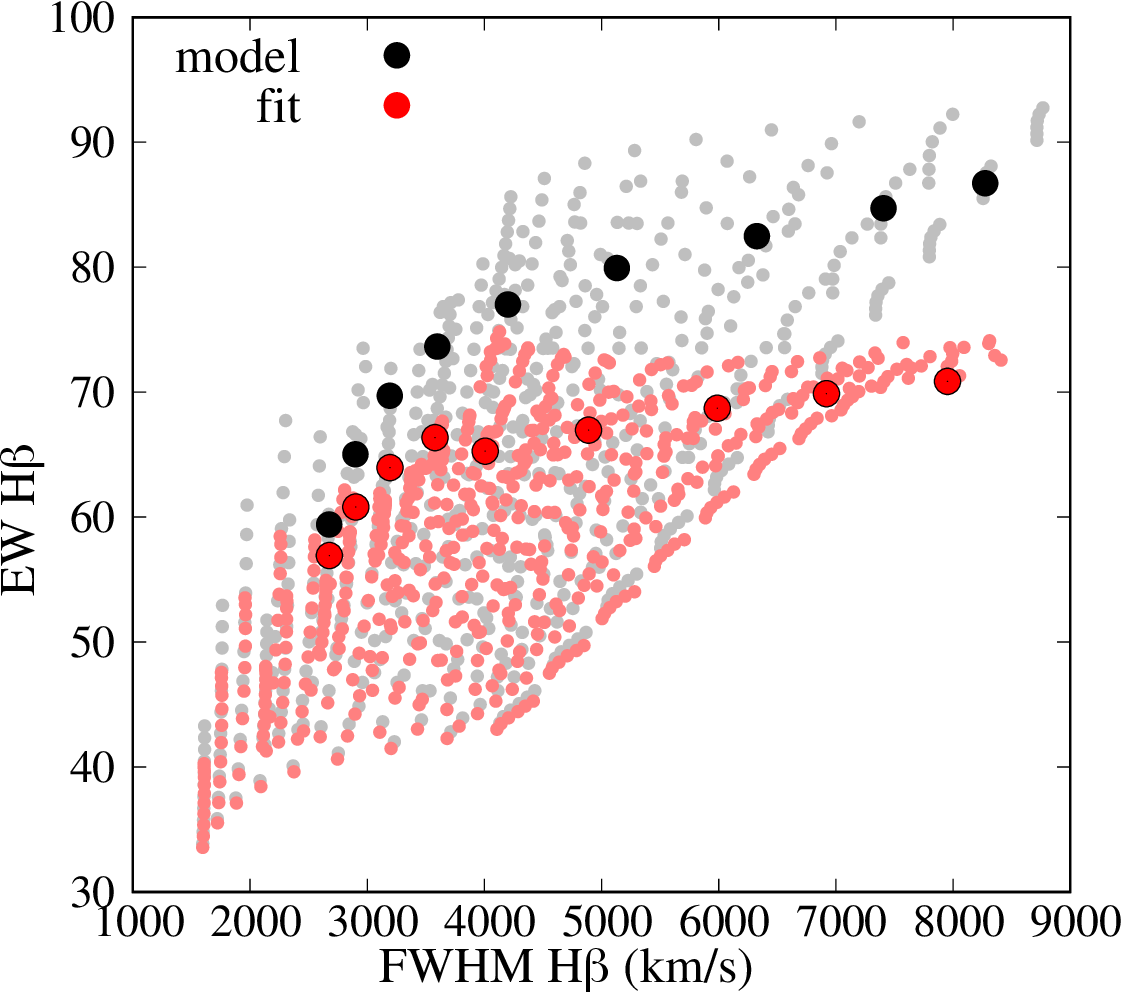}
\includegraphics[width=0.32\textwidth]{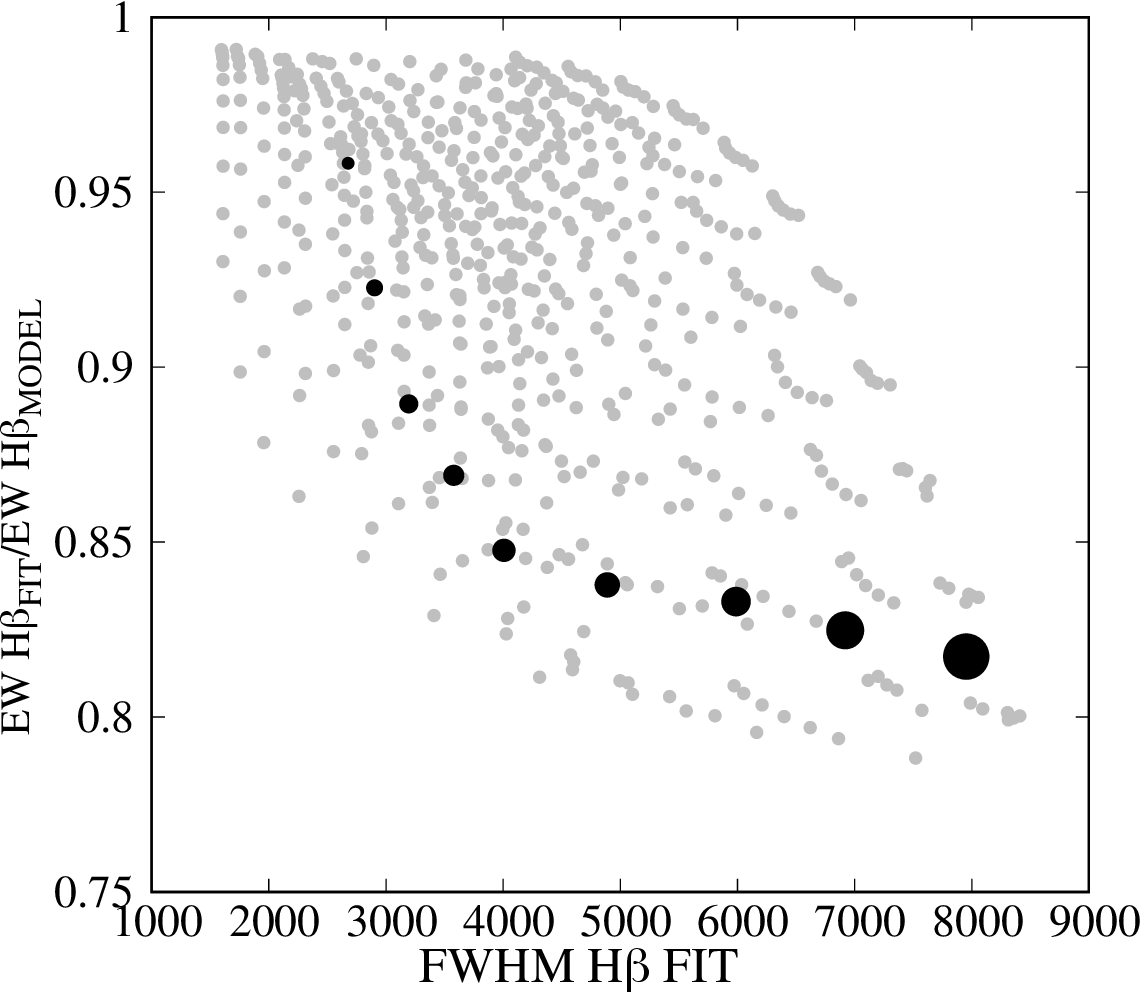}
\includegraphics[width=0.32\textwidth]{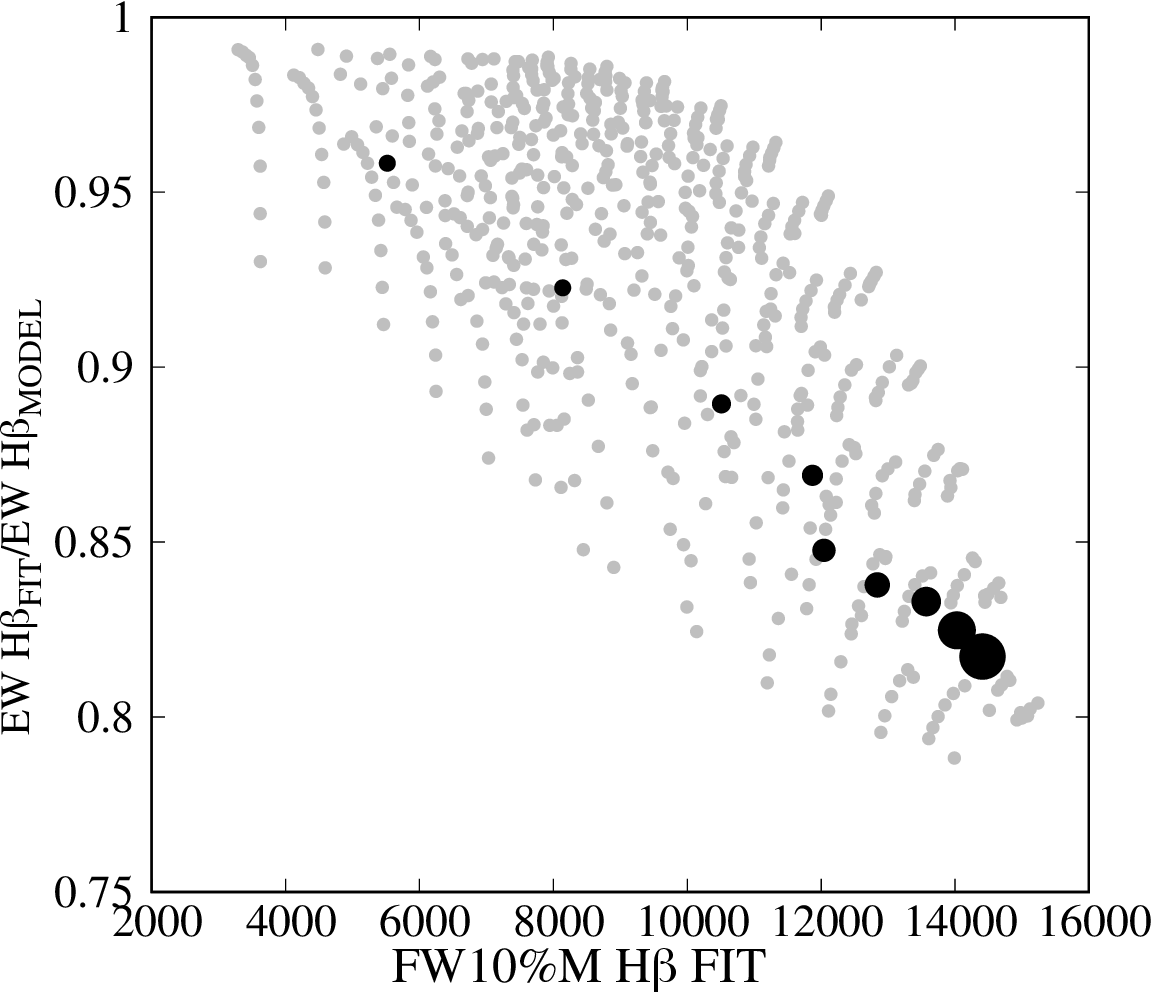}
\caption{Correlations between different spectral parameters. {\it Left:} Comparison between the EW H$\beta$ and FWHM H$\beta$ obtained from the best fit (red dots) and from the model (black dots) for the initial set of model spectra. The same is given for a large sample of model spectra, where parameters measured after fitting are assigned with light-red, small dots, and parameters from the model with grey dots. {\it Middle} and {\it Right}: Black dots are values obtained from the initial set of model spectra and their size increases with a larger VBLR contribution in the model spectra.
The grey dots are values obtained from the large sample of model spectra.
\label{fig6}}
 \end{figure*}
 
 \begin{figure*} 
 \centering
 \includegraphics[width=0.32\textwidth]{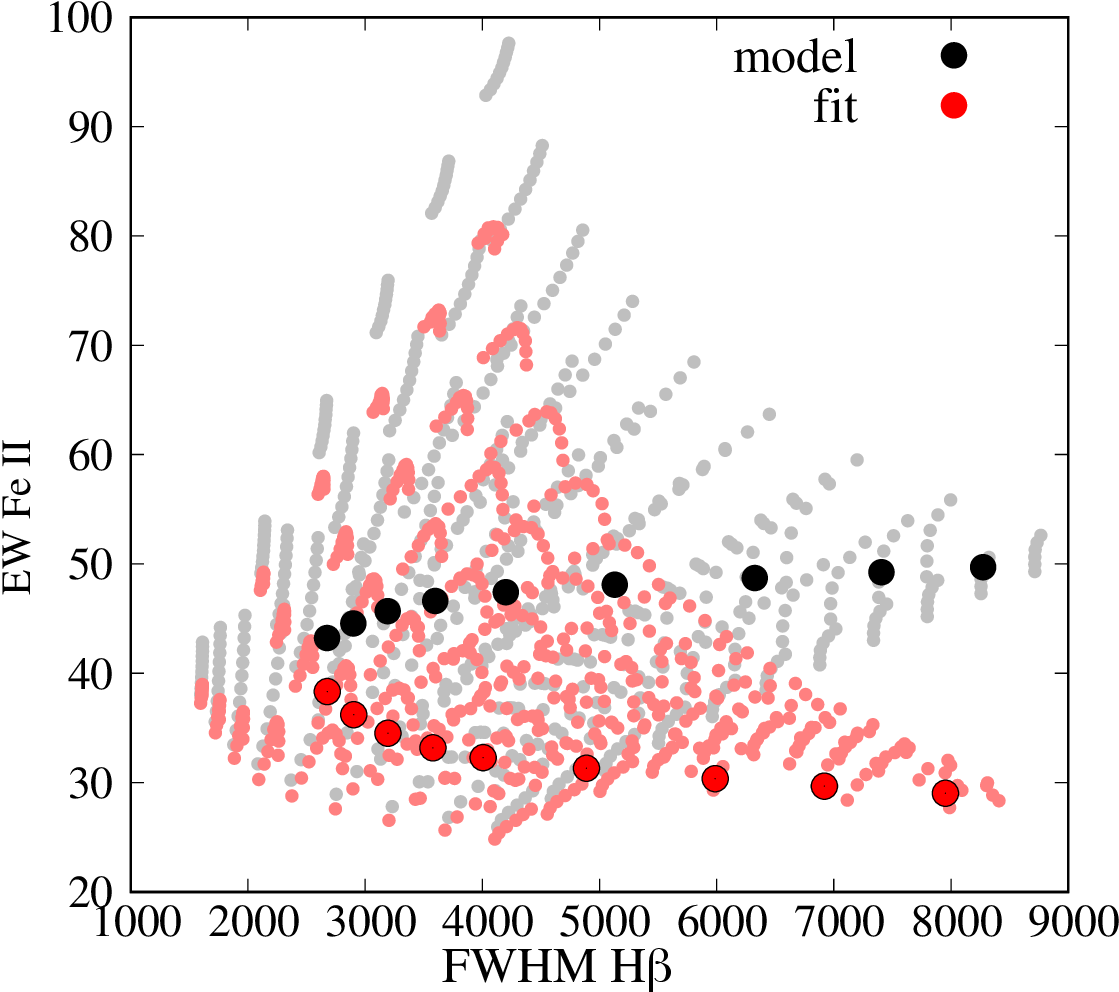}
\includegraphics[width=0.32\textwidth]{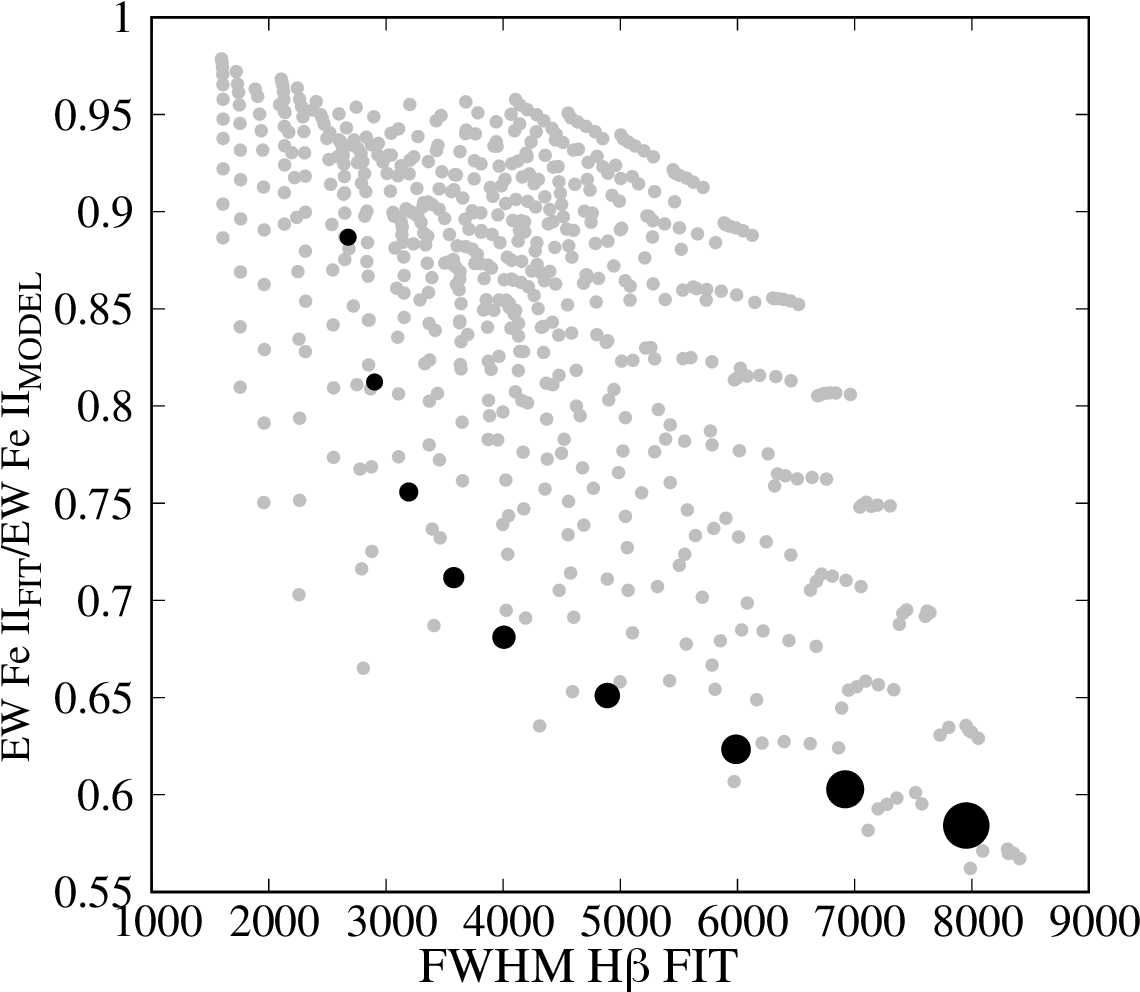}
\includegraphics[width=0.32\textwidth]{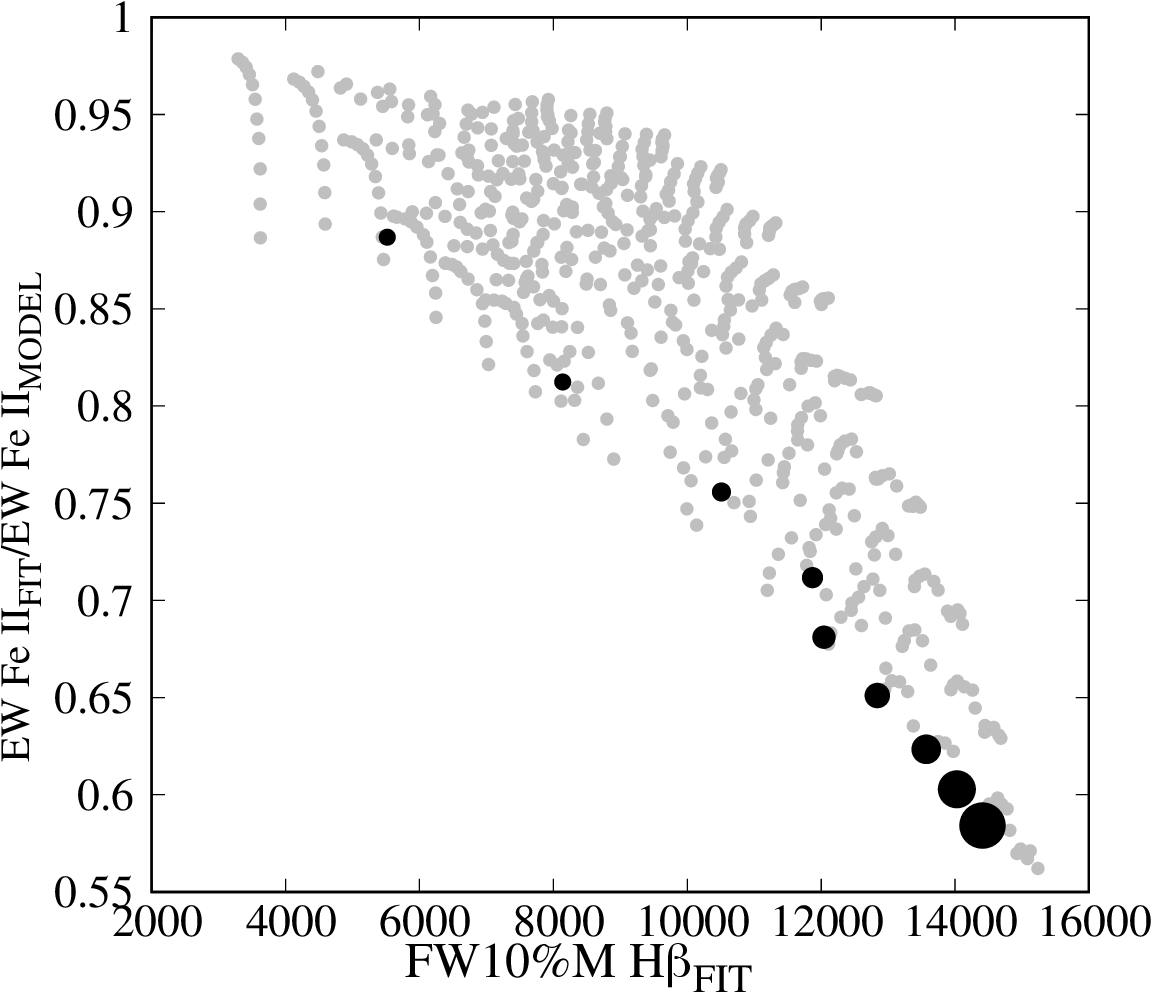}
\caption{Same as in Fig. \ref{fig6}, but for EW Fe II measured in the range of 4434-4684 \AA.
\label{fig6_1}}
 \end{figure*}

\subsection{Influence of the continuum subtraction on the measured spectral parameters}\label{3.2}

Here, we compare the spectral parameters obtained after continuum subtraction and the fitting procedure of the synthetic spectra, with the same parameters input in models. In Fig. \ref{fig6_0}, it can be seen that the H$\beta$ widths are not much affected by the procedure of continuum subtraction. The FWHM H$\beta$ is slightly underestimated (up to 10\%) comparing the model values in the case of the model spectra with very broad lines. Additionally, we measured the full width at 10\% of maximal intensity (FW10\%M) of H$\beta$ obtained from best fit and compared it with FW10\%M H$\beta$ of the model. The FW10\%M of H$\beta$ line is more affected by the process of continuum subtraction than the FWHM H$\beta$, since it can be underestimated when comparing the model up to 20\%. As expected, the width of H$\beta$ lines correlates with the VBLR contribution in model spectra, which is shown with different sizes of black points in Fig. \ref{fig6_0}.

The continuum flux at 5100 \AA \ obtained after the power-law fit is compared with the continuum flux of the model in Fig. \ref{fig6_00} (up). It can be seen that continuum flux obtained from the fit is slightly overestimated (up to 4\%) compared to the model and the disagreement is stronger for spectra with broader lines. Some studies follow the approximation that the spectrum at 5100 \AA \ is continuum window, with no line contribution at that wavelength, so the flux of the spectrum at 5100 \AA \ represents the flux of the continuum. In order to test the accuracy of that assumption, we directly measured the flux of the spectra at 5100 \AA \ and compared it with the flux of the continuum model (shown in Fig. \ref{fig6_00}, bottom). We found that the disagreement is larger than in the previous case, namely, the measured fluxes are overestimated by up to 10\% compared to the model values. The disagreement becomes more significant for spectra with broader lines.

The difference between the EW H$\beta$ measured after continuum subtraction and fitting procedure (EW H$\beta_{FIT}$) and the EW H$\beta$ measured directly from the model (EW H$\beta_{MODEL}$) is demonstrated in Fig. \ref{fig6}. The EWs of H$\beta$ obtained from fit are underestimated comparing the values from the model for up to 20\%. It can be seen that discrepancy is larger for broader  H$\beta$ lines and larger contribution of the VBLR (see the sizes of the black points). Also, it has a stronger correlation with FW10\%M H$\beta$ than with FWHM H$\beta$. Similarly to the case of EW H$\beta$, the difference between the EW Fe II (in the range of 4434-4684 \AA) measured after the continuum subtraction (EW Fe II$_{FIT}$) and EW Fe II measured directly from model (EW Fe II$_{MODEL}$) is shown in Fig. \ref{fig6_1}. The EWs of Fe II obtained from fit are more underestimated compared to the values from the model than the EWs of H$\beta$. The underestimation of the measured EW Fe II after fit procedure goes up to 45\% and it is the greatest for broader lines and for the spectra with the largest VBLR contribution. Similarly, as in the case of EW H$\beta$, the discrepancy correlates stronger with FW10\%M H$\beta$, than with FWHM H$\beta$.

\begin{figure} 
 \centering
  
\includegraphics[width=0.32\textwidth]{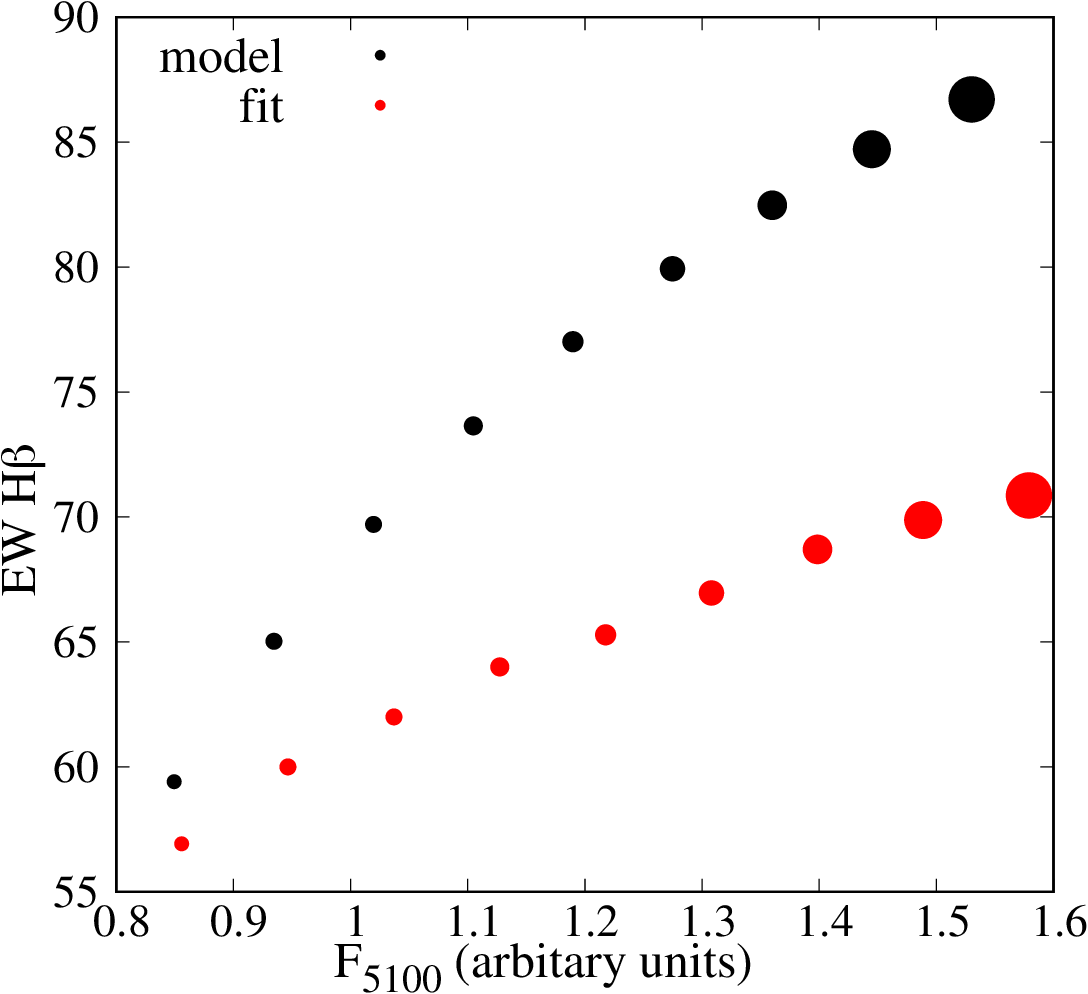} 
  \includegraphics[width=0.32\textwidth]{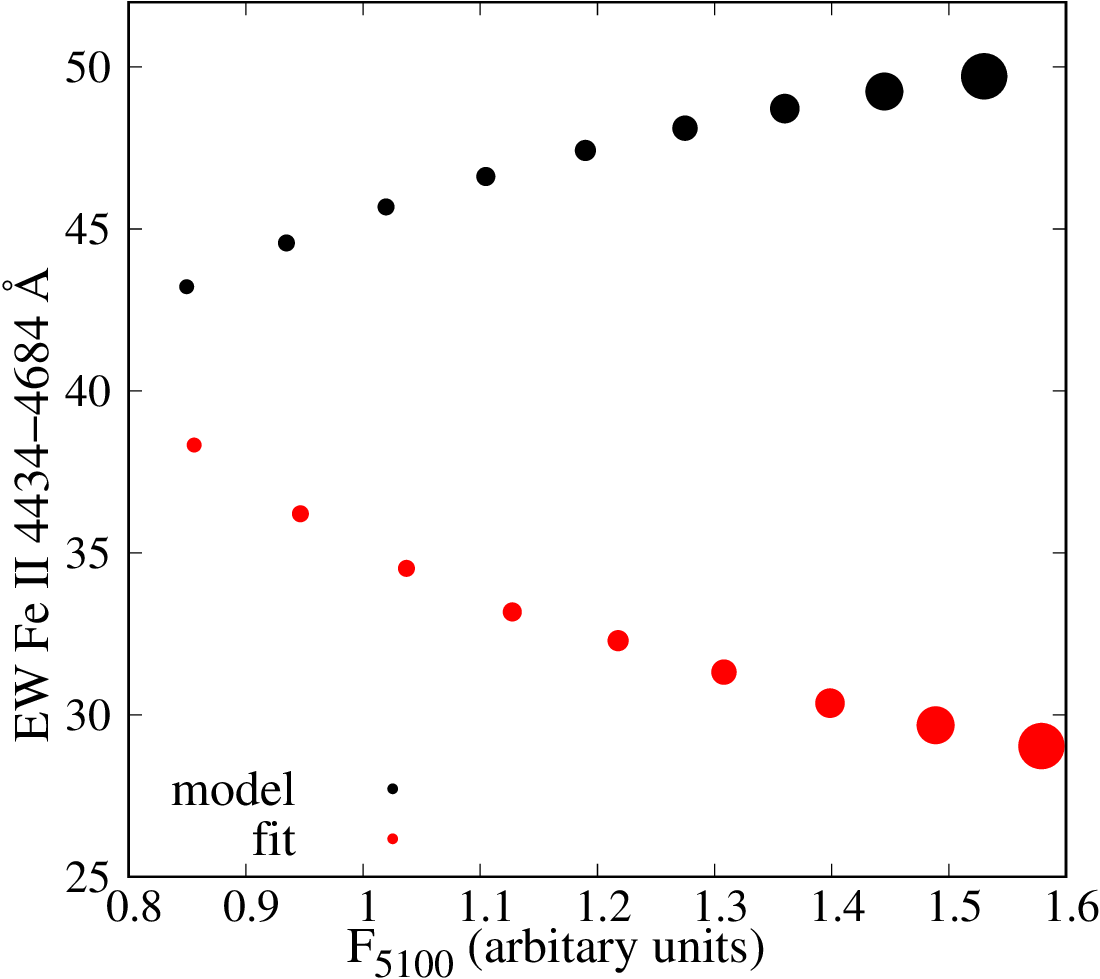} 
 \caption{ Correlations between line EWs and continuum flux. Correlation between EW H$\beta$ and continuum flux for parameters obtained from fit (red dots) and for parameters included in the model (black dots) for the initial set of model spectra {\it (top)}. The same just for EW Fe II 4434-4684 \AA \ and continuum flux {\it (bottom)}. The size of the dots denotes the contribution of the VBLR in the model spectra, where the largest dots represent the largest VBLR contribution.
 \label{fig7_4}}
 \end{figure}

The larger influence of continuum subtraction on the underestimation of EW Fe II compared to EW H$\beta$ could also be seen in Fig. \ref{fig7_4}, where we compare the EWs of these lines with the flux of the continuum at 5100 \AA. In the case of the initial set of models, the correlation EW Fe II versus Flux$_{5100}$ even becomes negative for measured EWs after fit. The reason for this is probably that part of the Fe II flux is joined to the continuum level during continuum fitting, so overestimation of the continuum and underestimation of the Fe II flux bring this trend.

\begin{figure*} 
 \centering

 \includegraphics[width=0.32\textwidth]{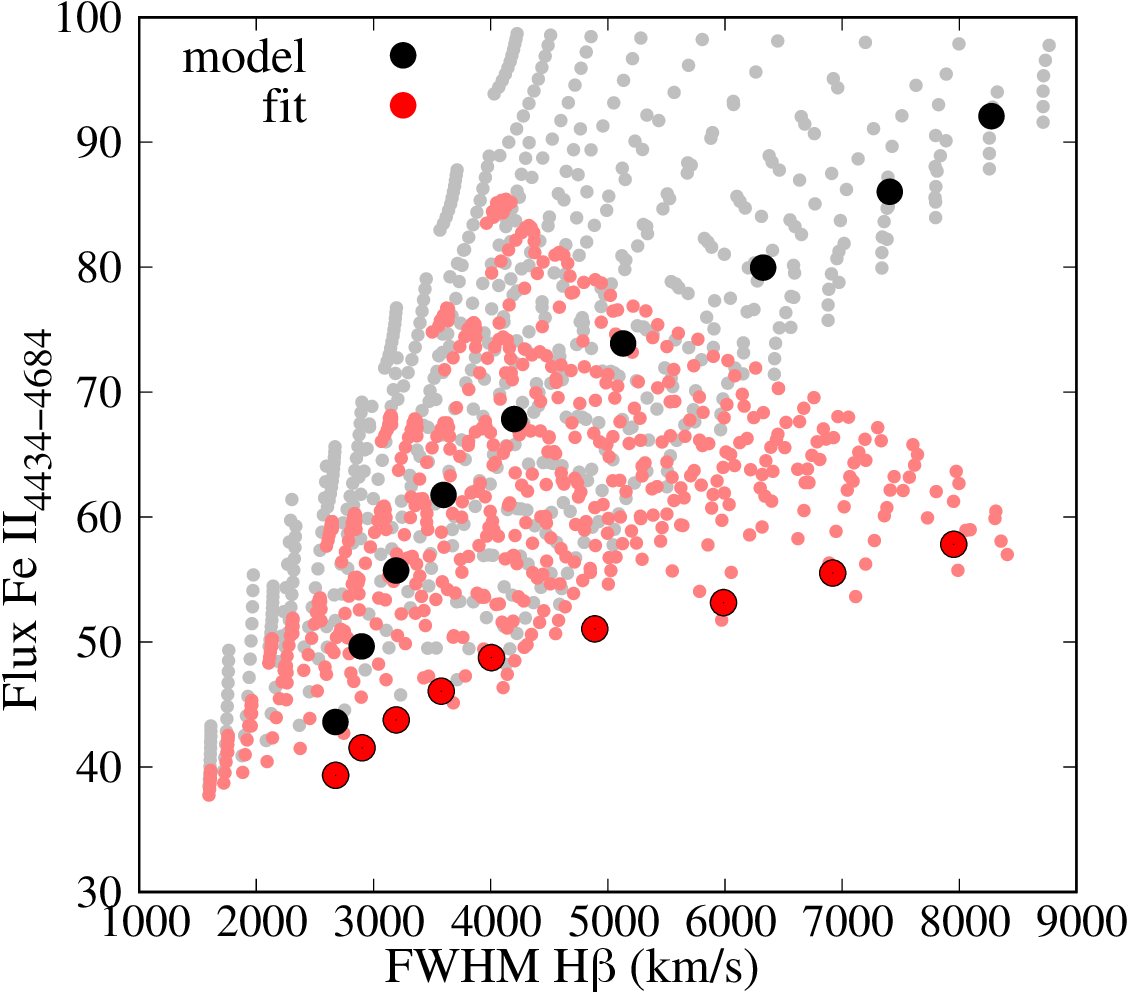} 
 \includegraphics[width=0.32\textwidth]{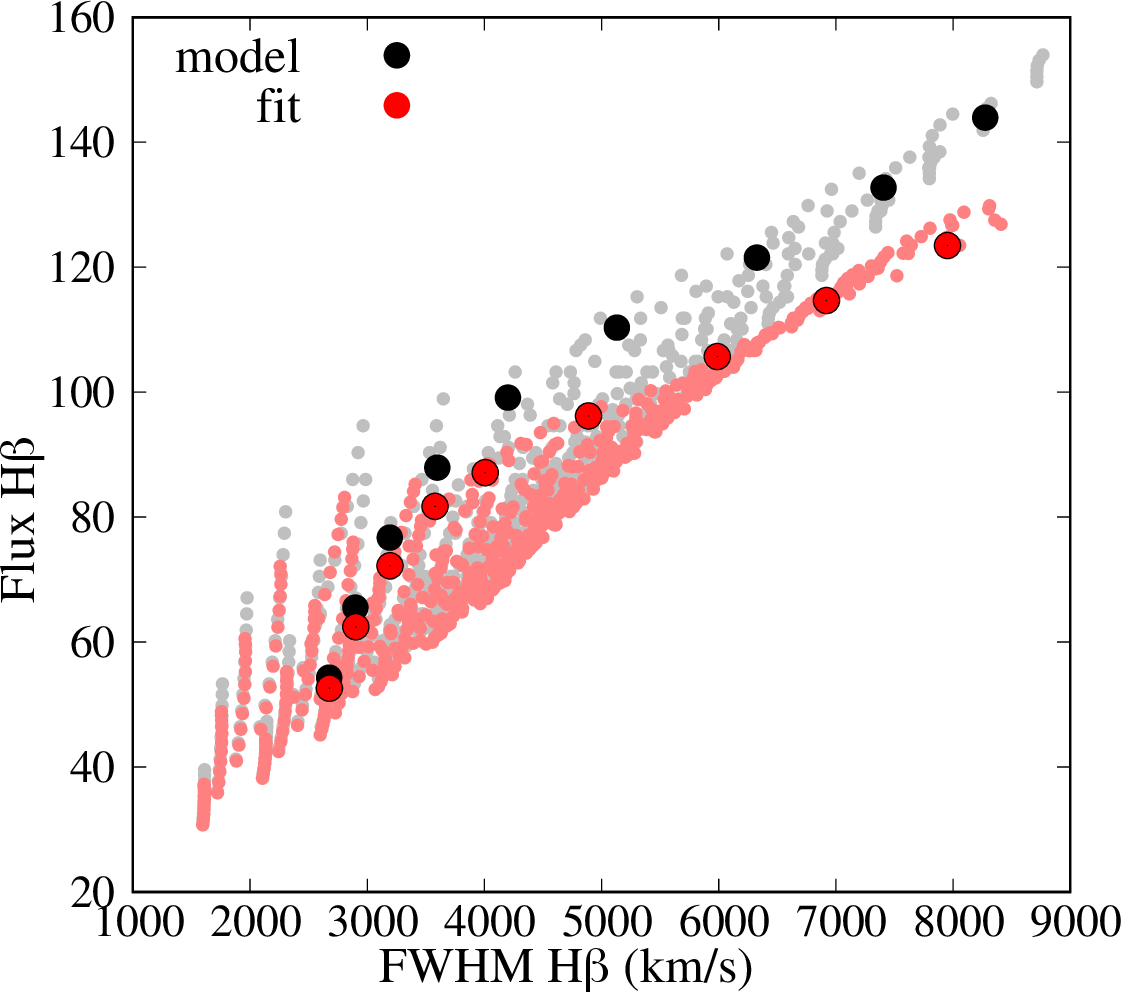}  
 \includegraphics[width=0.32\textwidth]{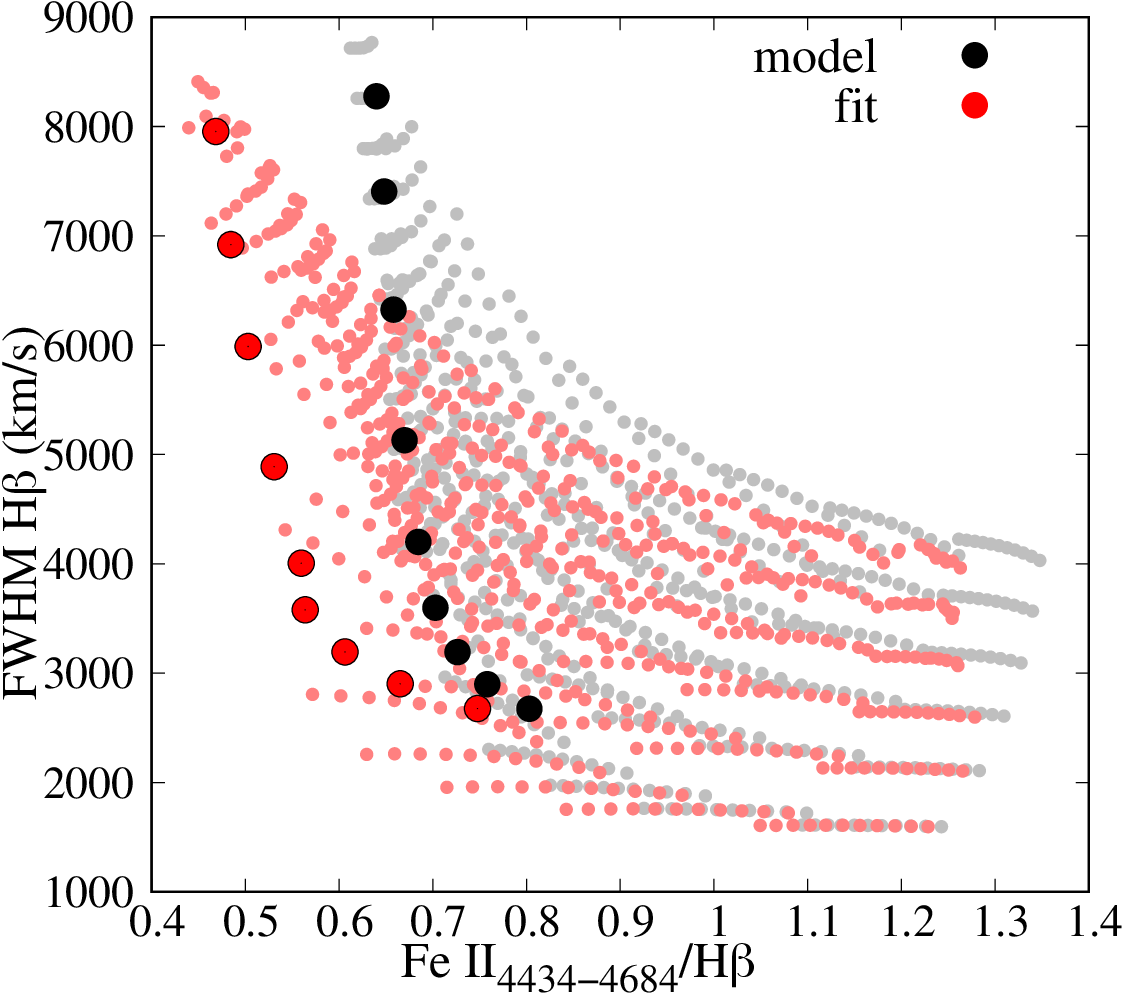} 
\caption{ Correlations between line fluxes versus FWHM H$\beta$, and FWHM H$\beta$ versus Fe II/H$\beta$ (EV1 correlation). The symbol denotation is the same as in Fig. \ref{fig6} (left).
\label{fig8}}
 \end{figure*}

We investigated the influence of the continuum subtraction on the fluxes of the Fe II in 4434-4684 \AA \ range and H$\beta$ lines in Fig. \ref{fig8} (left and middle). Similarly, as for EWs, the flux of the Fe II lines is significantly underestimated after continuum subtraction comparing model values, and difference becomes even larger for objects with broader lines. The flux of H$\beta$ lines remains similar to model values for FWHM H$\beta$ < 4000 km s$^{-1}$, while for spectra with broader lines, it starts to be slightly underestimated. We analysed the influence of the pseudo-continuum on the EV1 anti-correlation, FWHM H$\beta$ versus Fe II/H$\beta$, as shown in Fig. \ref{fig8} (right). The ratio of Fe II/H$\beta$ is underestimated for spectra with larger widths (up to 30\%), since the flux of Fe II is more affected by continuum subtraction than H$\beta$ flux, i.e., it contributes more to the quasi-continuum than H$\beta$.

 \subsection{Modelled and measured Eddington ratio and SMBH mass estimates}

 The H$\beta$ line and continuum at $\lambda$5100\AA\ are often used for the SMBH mass estimates. For the SMBH mass determination, using a virial approximation, we have to measure the FWHM H$\beta$ and luminosity at $\lambda$5100\AA,\ which indicates the dimension of the BLR \citep[see e.g. review] [and references therein]{pop20}. As can be seen in Fig. \ref{fig6_00}, the Fe II quasi-continuum affects the determination of the continuum level at $\lambda$5100 \AA, causing the overestimation of the continuum level and a slight underestimation of the FWHM H$\beta$ (see Fig. \ref{fig6_0}). Here, we investigate how it reflects on the estimates of the SMBH mass (M$_{BH}$) and Eddington ratio (R$_{Edd}$).

\begin{figure} 
 \centering
\includegraphics[width=0.32\textwidth]{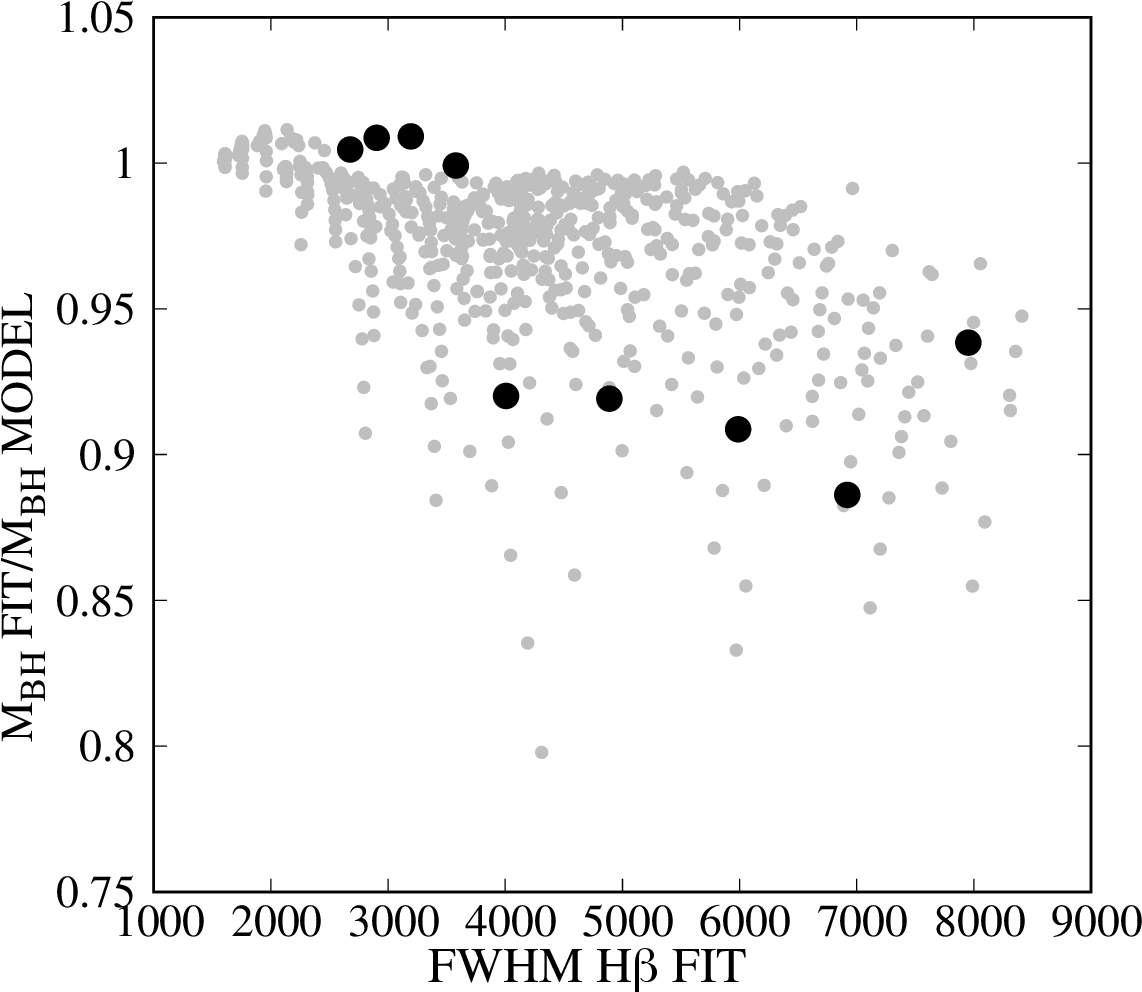} 
\includegraphics[width=0.32\textwidth]{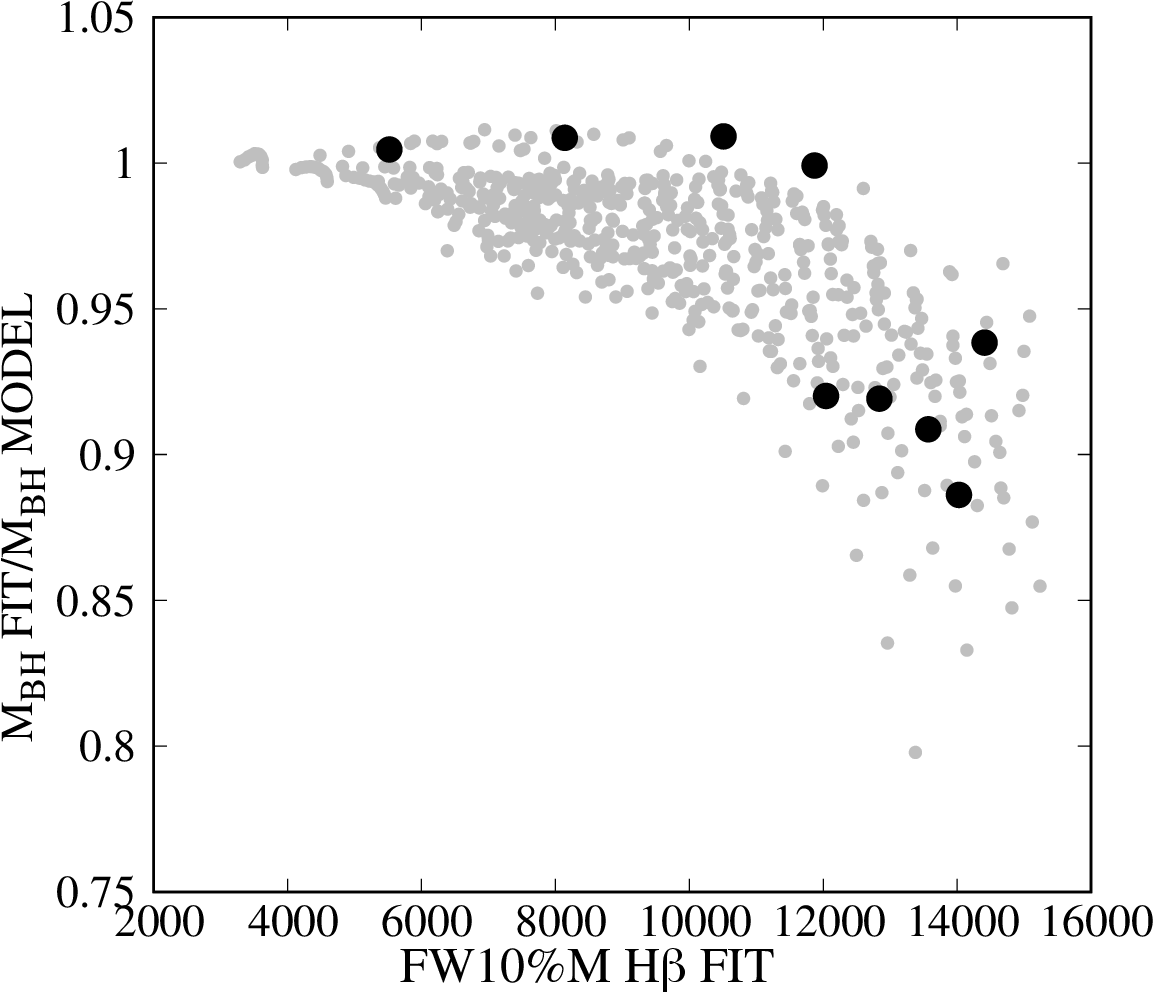} 
\caption{Comparison between M$_{BH}$ obtained using parameters measured 
after fitting procedure and M$_{BH}$ calculated with parameters from model, for spectra with different FWHM H$\beta$ {\it (top)} and FW10\%M H$\beta$ {\it (bottom)}. The symbol denotation is the same as in Fig. \ref{fig6_00}. 
\label{fig7_2}}
 \end{figure}
 
 \begin{figure} 
 \centering
 \includegraphics[width=0.32\textwidth]{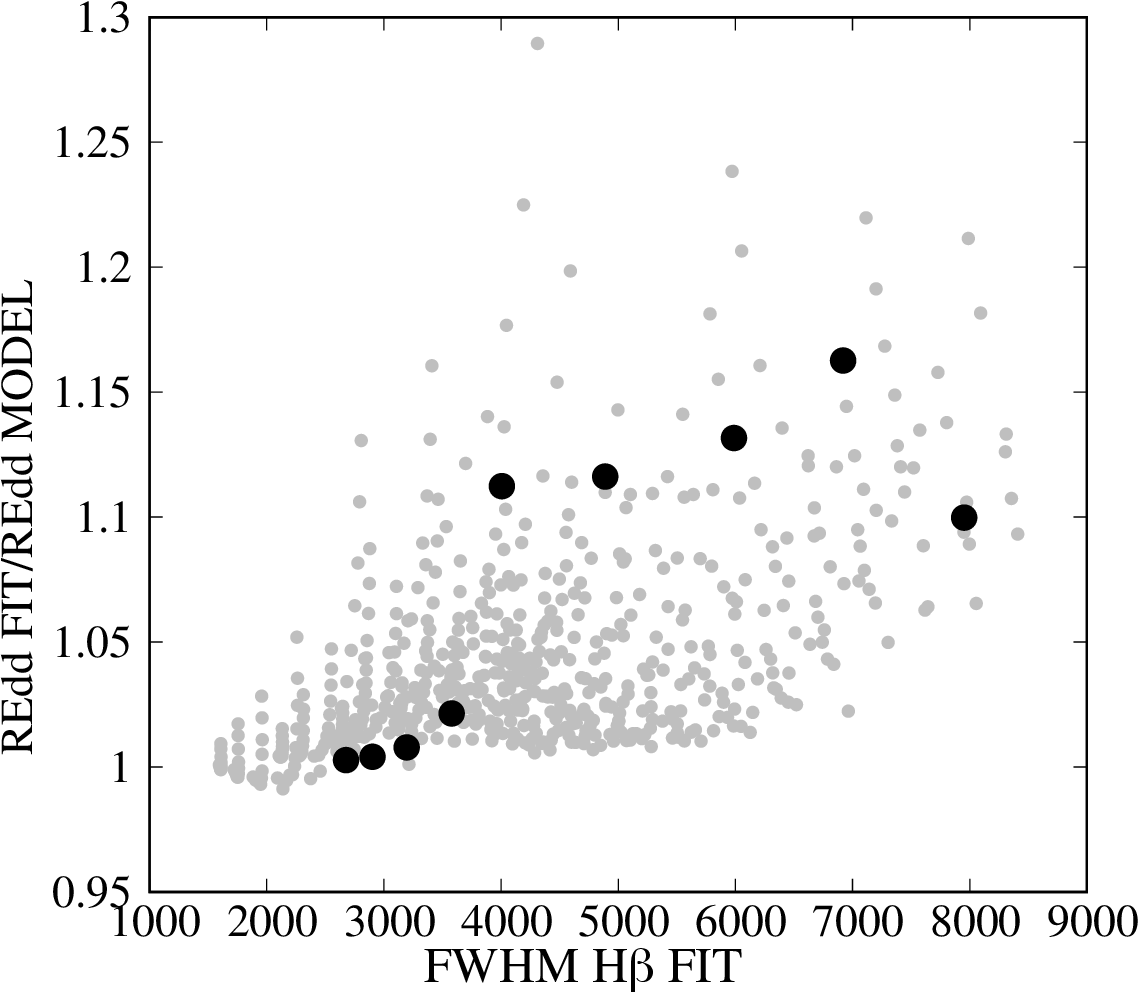} 
  \includegraphics[width=0.32\textwidth]{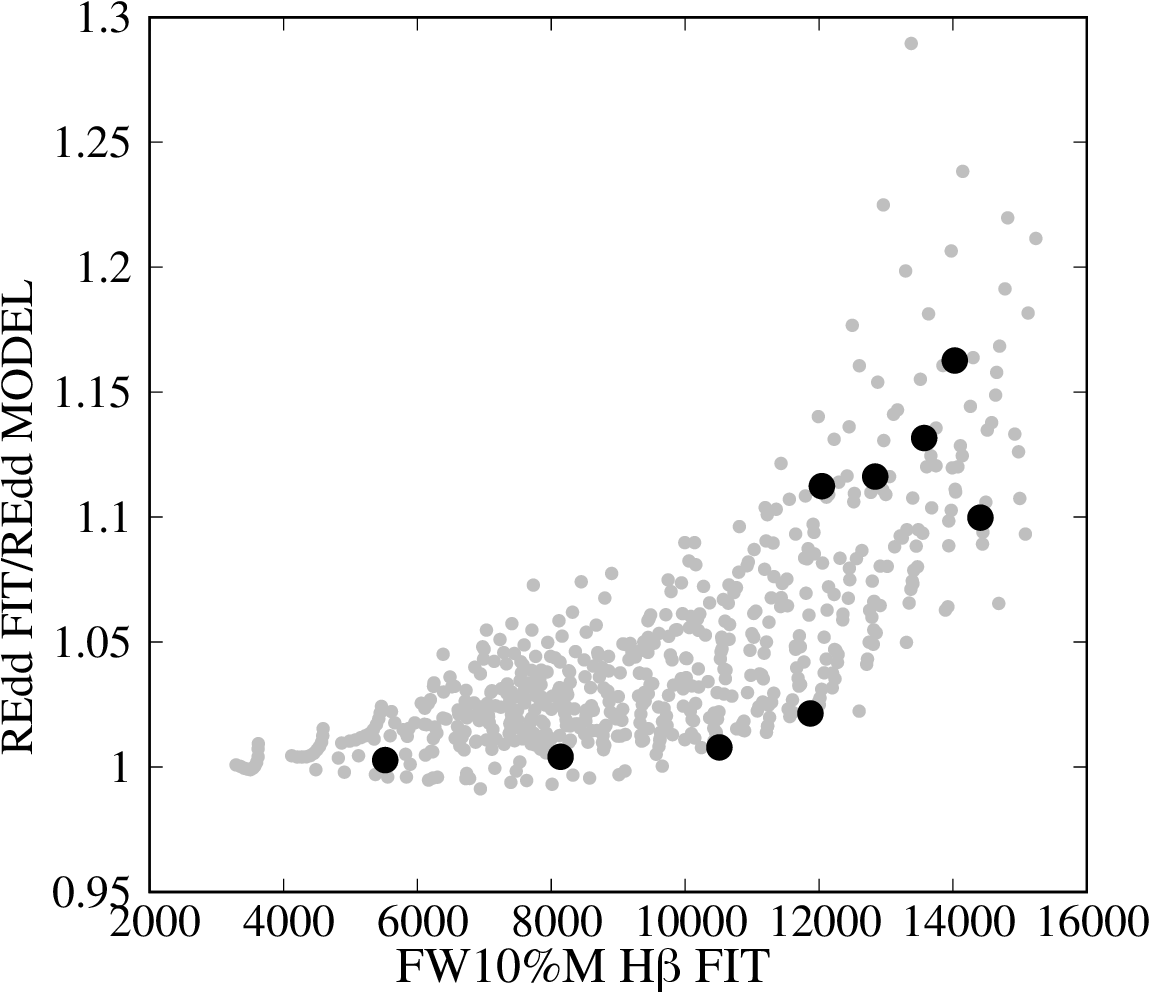} 
\caption{Same as in Fig. \ref{fig7_2}, but for the Eddington ratio, R$_{Edd}$. The symbol notation is the same as in Fig. \ref{fig6_00}. 
\label{fig7_3}}
 \end{figure}

 To explore the influence of the Fe II quasi-continuum on SMBH mass estimates, we used a common relationship for the mass determination in the case of virialisation:
$$ M_{BH}\sim FWHM\ H\beta^2\cdot R_{BLR},$$
where we assume that 
$$R_{BLR}\sim L^{0.5}_{5100}.$$
The $R_{BLR}$ is the broad line region radius and $L_{5100}$ is the continuum luminosity at $\lambda$5100 \AA. For our sample of synthetic spectra,  $L_{5100}$ is proportional to $F_{5100}$.
The ratio of the M$_{BH}$ obtained with measured parameters after the fitting procedure (M$_{BH\ FIT}$) with M$_{BH}$ obtained with input parameters in the model (M$_{BH\ MODEL}$) is calculated as:
$$ {M_{BH\ FIT}\over M_{BH\ MODEL}}= {FWHM\ H\beta_{FIT}^2\over {FWHM\ H\beta_{MODEL}^2}}{F_{5100\ FIT}^{0.5}\over {F_{5100\ MODEL}^{0.5}}}, $$
where the index {\it MODEL} denotes parameters included in the model and {\it FIT} denotes parameters measured after the fitting procedure.
We assume that $R_{Edd}\sim L_{5100}/M_{BH}$ \citep{Wu2004}. The R$_{Edd\ FIT}$/R$_{Edd\ MODEL}$ is calculated as:

$${R_{Edd\ FIT}\over R_{Edd\ MODEL}}={F_{5100\ FIT}\cdot M_{BH\ MODEL}\over M_{BH\ FIT}\cdot F_{5100\ MODEL}}.$$

The plots of ${M_{BH\ FIT}}/ {M_{BH\ MODEL}}$ ratio versus H$\beta$ width (Fig. \ref{fig7_2}) show that  M$_{BH}$ calculated using parameters obtained from fit is underestimated for spectra with broader lines, up to 15\%. On the other hand, R$_{Edd}$ calculated with parameters from fit is overestimated up to 25\% for spectra with broad lines (see Fig. \ref{fig7_3}). In both cases, correlations with H$\beta$ width become stronger for FW10\%M H$\beta$ (Fig. \ref{fig7_2} and Fig. \ref{fig7_3}, bottom).

\section{Discussion}\label{5.3}

Here, we discuss the possible underlying physics in the quasar main sequence implied by the results of our spectral modelling.
Then, we draw attention to the possible influence of the Fe II quasi-continuum on the measured spectral parameters and its implications for drawing conclusions about AGN physics.
 
 \subsection{Quasar main sequence:\ The nature}

The so-called quasar main sequence, which represents the anticorrelation of the FWHM H$\beta$ versus optical Fe II line strength relative to H$\beta$, indicates physical differences between various AGNs \citep[see e.g.][]{Boroson1992,su00,sh14,ma18}. This correlation divides AGNs into two subgroups, Pop A and Pop B \citep[see][]{ma18}, where Pop A are objects with FWHM H$\beta$ < 4000 km s$^{-1}$ and a high accretion rate, while Pop B are objects with FWHM H$\beta$ > 4000 km s$^{-1}$ and a lower accretion rate. 

The complex shape of the broad H$\beta$ lines observed in various AGN spectra implies that these lines arise in two subregions of the BLR  (VBLR and ILR), closer and further away from the SMBH \citep{bo09}. On the other hand, it is commonly considered that Fe II emission arises in the same region that produces the broad H$\beta$ line \citep{Phillips1978,Boroson1992}. In this research, we assume that both lines, Fe II and H$\beta$, have two components: one very broad (VBLR) and one intermediate (ILR), both coming from the BLR, but from different layers. Using this assumption, we constructed a set of model spectra with different contributions of ILR and VBLR emission in H$\beta$ and Fe II optical lines, where the relative intensities between Fe II and H$\beta$ lines are adopted from real, observed spectra identified as dominantly ILR and dominantly VBLR emission (see Figs. \ref{fig1} and \ref{fig2}). The adopted Fe II/H$\beta$ ratio is larger in the ILR dominant prototype spectrum than in the VBLR dominant spectrum. The obtained set of the model spectra has parameters FWHM H$\beta$ and FeII/H$\beta$ in an empirically expected range (see Fig. \ref{fig5_11}), within the contour obtained using the large SDSS AGN sample \citep{sh14}. It is interesting to note that this set of model spectra reproduced the expected anticorrelation FWHM H$\beta$ versus FeII/H$\beta$. Moreover, we found that this anticorrelation is strongly dependent on the rate of the VBLR and ILR contributions in the model spectra (Fig. \ref{fig5_1}). In our set of model spectra, the models with VBLR contribution in the range of 0.1-0.5 mostly have FWHM H$\beta$ < 4000 km s$^{-1}$, that is, they belong to Pop A, while the models with VBLR contribution in the range of 0.5-0.9 mostly have FWHM H$\beta$ > 4000 km s$^{-1}$, that is, they belong to Pop B.

This result opens up two questions:\ first, we consider what causes a different ratio of Fe II/H$\beta$ in the ILR and VBLR regions; second, what controls the different rate of the ILR and VBLR contributions in various AGN spectra.
 
The model of locally optimally emitting clouds (LOC), given in \cite{Baldwin1995}, predicts that the lines predominantly arise in regions in which physical conditions for their emission are optimal. Therefore, it is possible that H$\beta$ and Fe II lines dominantly arise from two LOCs, ILR, and VBLR. Since the VBLR is closer to SMBH than ILR, it is possible that different physical conditions in these two regions affect the atomic processes that might produce Fe II lines, such as radiative excitation, collisional excitation, fluorescent excitation by Ly$\alpha$, and so on, leading to their different efficiency in ILR and VBLR.

 On the other hand, it is commonly accepted that Fe II and H$\beta$ emission lines are dominantly produced by photoionisation \citep[see][and references therein]{ga22}. Namely, on the front side of the gas clouds, the hydrogen would be ionised, and H$\beta$ would be produced in the process of the recombination. The Fe II emission would arise dominantly from the ionisation front at the farther side of clouds, where hydrogen is mostly neutral, in the process of collisional excitation \citep{ga09,ga22}. In this scenario, the Fe II emission could only be produced in gas clouds thick enough for an ionisation front to happen but also with a large enough column density. In clouds of lower column density, only H$\beta$ would be produced, but no Fe II emission. The VBLR could be made of a range of clouds of different thicknesses, all producing H$\beta,$ but with only the thicker one (i.e. high column density) would produce Fe II \citep{jo1987}. This results in a smaller  Fe II/H$\beta$ intensity ratio emitted from VBLR. On the other hand, in the farther away ILR, the ionisation parameter would be smaller and Fe II emission would be higher \citep{wi1985}.  Thus, the different physical conditions in the VBLR and ILR regions might lead to different Fe II/H$\beta$ intensity ratios. The observed  spectrum is then the sum of ILR and VBLR contributions in emission lines, which gives the quasar main sequence, namely:\ the EV1 parameter space. The increase in the VBLR contribution in Fe II lines consequently increases the width of the lines, but decreases the Fe II/H$\beta$ ratio. It is in accordance with the state that ionisation parameters increase from Pop A to Pop B \citep{ma2001}.

However, it is an interesting question what controls the rate of the VBLR/ILR contribution in the AGN spectra, especially in the Fe II and Balmer lines. We propose that the VBLR/ILR contribution is controlled by both inclination and accretion rate, similarly to what was noticed in \cite{sh14} who found that the main sequence is controlled by these parameters. The dominant ILR emission in the observed spectra can be caused by a large inclination, that is, the VBLR component could be obscured. On the other hand, a high level of accretion might produce the physical conditions of large radiation pressure, which prevent the formation of the Fe II emission (and also H$\beta$) close to the central SMBH, so the line-emitting region is formed farther away from the central SMBH. In that case, we can observe narrower Fe II emission lines. In the case where the VBLR is dominant in the spectra, the accretion rate is not very high, resulting in lower radiation pressure, since the emission is predominantly coming from the region that is close to the central SMBH. Also, the inclination should be such that we can see the dipper in the centre of an AGN.

 \subsection{Optical Fe II quasi-continuum in AGN spectra}

In the two-component BLR model, the VBLR component contributes to the broad line wings, while the ILR component dominates the line core. In the case of the spectra with a dominant VBLR component, an nearly complete broad H$\beta$ profile could be well observed, but the broad wings of Fe II lines overlap and form the Fe II quasi-continuum, making it difficult to separate it from the real continuum. In the case of the ILR-dominant spectra, the VBLR components in Fe II are too weak to have a significant contribution to the Fe II quasi-continuum.

In analysing the set of model spectra, our results imply that Fe II quasi-continuum affects the process of line and continuum fitting and estimation of the spectral parameters. In the spectra with dominant VBLR contribution, the width of the H$\beta$ is slightly underestimated, especially when measured at the level of the H$\beta$ wings (FW10\%M). On the other hand, the estimated continuum flux is slightly overestimated since part of the line flux is included. Generally, the measured properties of the H$\beta$ lines, such as line flux and EW, are less affected by the process of continuum subtraction than the flux and EW of Fe II, which implies that Fe II lines contribute much more to the quasi-continuum than the broad H$\beta$ line. The fluxes of the emission lines are underestimated, but line EWs are the most sensitive to the influence of the quasi-continuum, which is a consequence of the flux underestimation and continuum overestimation. In general, it seems that the spectral parameter that is most sensitive to quasi-continuum presence is the EW Fe II. Therefore, we should be very careful with measured spectral parameters in the case of spectra with strong and broad Fe II lines, especially with respect to the EW of Fe II.

Some important AGN properties, such as M$_{BH}$ and R$_{Edd}$, which are calculated using the measured spectral parameters of the FWHM H$\beta$ and continuum luminosity, are consequently affected by the Fe II quasi-continuum contribution. For spectra with strong and broad Fe II lines, the M$_{BH}$ could be slightly underestimated and R$_{Edd}$ could be overestimated, which should be taken into account.

The influence of the Fe II quasi-continuum on measured parameters may affect some conclusions obtained from AGN variability, such as the measured dimension of the Fe II emission region by reverberation. Namely, if the Fe II quasi-continuum is incorporated in measured continuum flux used in reverberation mapping technique, it would cause uncertainty of the measured time lags for the Fe II emission lines,  namely, it would unable its measuring. We expect that this effect would be specially present in spectra with broad and strong Fe II lines in which Fe II quasi-continuum significantly contributes to the real continuum. It is interesting to note that Fe II time lags are measured in only about 10\% of AGNs for which the reverberation mapping was applied; whereas in the other studies, the Fe II time lags could not be measured properly. For example, \cite{Kuehn2008} carried out a reverberation mapping for Ark 120, which is an object with broad and strong Fe II lines, and they were not able to measure a clear time lag for the Fe II lines. 
 
Additionally, \cite{hu15} found a tentative correlation between the Fe II time lag and intensity ratio of Fe II and H$\beta$ that might be affected by Fe II quasi-continuum. \cite{du19} found that R$_{BLR}$ - L$_{5100}$ relation \citep[see][]{Bentz2013} should be corrected for a factor that includes the Fe II/H$\beta$ intensity ratio (see their Fig. 5 and Eq. 5) that may also be influenced by the Fe II quasi-continuum.

\section{Conclusions}\label{6}

We modelled the optical spectral region around H$\beta$ and Fe II optical emission, which is frequently used to point out some spectral characteristics of AGNs. We considered the fact that the Fe II and H$\beta$ emission regions have two components: one very broad (VBLR) and one intermediate (ILR), both coming from BLR. Using this assumption, we constructed the set of synthetic spectra with different contributions of ILR and VBLR emission in  H$\beta$ and Fe II optical lines, using a linear combination of the ILR and VBLR model spectra. The VBLR model spectrum has a significantly larger width and a smaller Fe II/H$\beta$ ratio compared to the ILR model spectrum, as taken from the two real, observed spectra with dominant VBLR, namely, the ILR emission. We analysed the obtained set of model spectra in context of the quasar main sequence and the influence of the ILR/VBLR contribution on the EV1 parameter space. In order to investigate the influence of the Fe II pseudo-continuum on the measured parameters in AGN spectra, we subtracted the continuum emission from model spectra by fitting the power law. Then we fitt the Fe II and Balmer lines. In this way, we were able to compare the parameters obtained after continuum subtraction and fitting procedures with those included in the model.
Summarising the main results, we can outline the following conclusions.

\begin{enumerate} [(i)]
 \item The observed broad-line AGN spectra in optical could be aptly modelled as the sum of the emission from two regions, ILR and VBLR, where each of these two regions emits spectra with different physical properties, reflecting the specific physical conditions in these regions.

 \item The set of model spectra constructed as different rates of the ILR/VBLR contribution could reproduce the quasar main sequence, namely, FWHM H$\beta$ versus Fe II/H$\beta$ anticorrelation. Both parameters in this anticorrelation are strongly dependent on the ILR/VBLR rate.

 \item The quasar main sequence is probably controlled by several effects: a) physical conditions in ILR and VBLR regions, which could cause different efficiency of some of atomic processes that produce H$\beta$ and Fe II lines, as well as different ionisation parameters in thin and thick clouds, making the Fe II/H$\beta$ ratio larger in ILR than in VBLR; b) the inclination and Eddington ratio, which control the  contribution of the ILR or VBLR emission in each spectrum.

\item In the case of the strong VBLR contribution, the VBLR components of the Balmer lines and Fe II lines make up the quasi-continuum in AGN spectra, where the Fe II quasi-continuum strongly dominates over the H$\beta$ quasi-continuum.

\item The Fe II quasi-continuum is strong in the case of the broad and strong Fe II lines and it is difficult to separate it from the real continuum. It may affect the process of the continuum and line fitting, causing an underestimation of the width of H$\beta$ and especially the Fe II and H$\beta$ fluxes, as well as an overestimation of the measured continuum level. The most affected spectral parameters are line EWs, especially EW Fe II, which may be strongly underestimated.

\item  Some estimated AGN properties could be affected by Fe II quasi-continuum in extreme cases, such as M$_{BH}$, as it may be underestimated, and R$_{Edd}$, which may be overestimated. Also, it is  important to be careful with clues obtained with the reverberation mapping technique, in the case of strong and broad Fe II lines, which may partly contribute to the measured continuum and may affect the results in that way.
   
\end{enumerate}

\begin{acknowledgements}
This work is supported by the Ministry of Education, Science and Technological Development of Serbia (451-03-47/2023-01/200002 and 451-03-47/2023-01/200162). L. \v C. Popovi\'c thanks the support by Chinese Academy of Sciences President’s International Fellowship Initiative (PIFI) for visiting scientist. 

\end{acknowledgements}

%-------------------------------------------------------------------


\begin{thebibliography}{}

\bibitem[Baldwin et al. (1995)]{Baldwin1995} Baldwin J., Ferland G., Korista K., Verner D. 1995, ApJ, 455, L119

\bibitem[Barth et al. (2013)]{ba13} Barth, A. J., Pancoast, A., Bennert, V.  N. et al. 2013, ApJ, 769, 128

\bibitem[Bentz et al. (2013)]{Bentz2013} Bentz, M. C., Denney, K. D., Grier, C. J. et al. 2013, ApJ, 767, 149

\bibitem[Bon et al. (2009)]{bo09} Bon, E. Popovi\'c, L. \v C., Gavrilovi\'c, N., La Mura, G., Mediavilla, E.
 2009, MNRAS, 400, 924 

\bibitem[Bon et al. (2006)]{bo06} Bon, E., Popovi\'c, L. \v C., Ili\'c, D., Mediavilla, E.
2006, NewAR, 50, 716 

\bibitem[Boroson \& Green (1992)]{Boroson1992} Boroson, T. A., Green, R. F., 1992, \apjs, 80, 109

\bibitem[Collin et al. (2006)]
{co06} Collin, S.,  Kawaguchi, T.,  Peterson, B. M.,  Vestergaard, M. 2006, A\&A, 456, 75

\bibitem[Collin-Souffrin et al. (1980)]{co80} 
Collin-Souffrin, S., Dumont, S., Heidmann, N., Joly, M. 1980, A\&A, 83, 190 


\bibitem[Dong et al. (2008)]{Dong2008} Dong X., Wang T., Wang J., Yuan W., Zhou H., Dai H., Zhang K. 2008, MNRAS, 383, 581

\bibitem[Du \& Wang (2019)]{du19}
Du, P. Wang, J.-M. 2019, ApJ, 886, 42


\bibitem[Gaskell (2009)]{ga09} Gaskell, C. M. 2009, NewAR, 53, 140


\bibitem[Gaskell et al. (2022)]{ga22} Gaskell, C. M., Thakur, N., Tian, B., Saravanan, A. 2022, AN, 34310112 


\bibitem[Hu et al. (2015)]{hu15}
Hu, C., Du, P., Lu, K.-X. et al. 2015, ApJ, 804, 138


\bibitem[Hu et al. (2020)]{hu20} Hu, C., Li, S.-S., Guo, W.-J.  et al. 
2020. ApJ, 905, 75 

\bibitem[Hu et al. (2012)]{hu12} Hu, C., Wang, J.-M., Ho, L. C., Ferland, G. J., Baldwin, J. A., Wang, Y. 2012, ApJ, 760, 126 

\bibitem[Jarvis \& McLure (2006)]{jm06} Jarvis, M. J., McLure, R. J. 2006, MNRAS, 369, 182

\bibitem[Joly (1981)]{jo81} Joly, M. 1981, A\&A, 102, 321 

\bibitem[Joly (1987)]{jo1987} Joly, M. 1987, A\&A, 184, 33 

\bibitem[Kova\v{c}evi\'{c}-Doj\v{c}inovi\'{c} \& Popovi\'{c} (2015)]{kovacevic2015} Kova\v{c}evi\'{c}-Doj\v{c}inovi\'{c}, J., Popovi\'{c}, L. \v C., 2015, \apjs, 221, 35


\bibitem[Kova\v{c}evi\'{c} et al.(2010)]{kovacevic2010} Kova\v{c}evi\'{c}, J., Popovi\'{c}, L. \v C. \& Dimitrijevi\'{c}, M.,\ 2010, \apjs, 189, 15

\bibitem[Kuehn et al. (2008)]{Kuehn2008} Kuehn, C.A., Baldwin, J.A., Peterson, B.M., Korista, K.T. 2008, ApJ, 673, 69 

\bibitem[Kuraszkiewicz et al.(2002)]{Kuraszkiewicz2002} Kuraszkiewicz, J. K., Green, P. J., Forster, K., Aldcroft, T. L., Evans, I. N., \& Koratkar, A. 2002, ApJS, 143, 257
 
 \bibitem[Laki\'cevi\'c et al. (2022)]{la22}
Laki\'cevi\'c, M.,  Kova\v{c}evi\'{c}-Doj\v{c}inovi\'{c}, J., Popovi\'c, L. \v C. 2022, MNRAS, 509, 831

\bibitem[Laki\'cevi\'c et al. (2018)]{la18} Laki\'cevi\'c, M, Popovi\'c, L. \v C., Kova\v{c}evi\'{c}-Doj\v{c}inovi\'{c}, J. 2018, MNRAS, 478, 4068
 
\bibitem[Le \& Woo (2019)] {le19} Le, H. A. N., Woo, J.-H. 2019, ApJ, 887, 236 

\bibitem[Marinello et al. (2016)]{ma16} Marinello, M., Rodríguez-Ardila, A., Garcia-Rissmann, A., Sigut, T. A. A., Pradhan, A. K. 2016, ApJ, 820, 116 


\bibitem[Marziani et al. (2021)]{ma21} Marziani, P., Berton, M., Panda, S., Bon, E. 2021, Univ, 7, 484 

\bibitem[Marziani et al. (2018)]{ma18} Marziani, P., Dultzin, D., Sulentic, J. W. et al. 2018, FrASS, 5, 6 

\bibitem[Marziani et al. (2001)]{ma2001} Marziani, P., Sulentic, J. W., Zwitter, T., Dultzin-Hacyan, D., Calvani, M. 2001, ApJ, 558, 553

\bibitem[Netzer (2013)]{ne13} Netzer, H. 2013, The Physics and Evolution of Active Galactic Nuclei, Cambridge, UK: Cambridge University Press

\bibitem[Osterbrock \& Ferland (2006)]{of06}  Osterbrock, Donald E.; Ferland, Gary J. 2006, Astrophysics of gaseous nebulae and active galactic nuclei,  CA: University Science Books
%

\bibitem[Park et al. (2022)]{pa22} Park, D., Barth, A. J., Ho, L. C., Laor, A. 2022, ApJS, 258, 38 

\bibitem[Peterson (1997)]{pe97} Peterson, B. M. 1997,  An introduction to active galactic nuclei, Cambridge, New York: Cambridge University Press

\bibitem[Phillips (1978)]{Phillips1978} Phillips M. M. 1978, ApJS, 38, 187

\bibitem[Popovi\'c (2020)]{pop20}
Popovi\'c, L. \v C. 2020, OAst, 29, 1

\bibitem[Popovi\'c \& Kova\v cevi\'c (2011)]{pk11}
Popovi\'c, L. \v C.,  Kova\v cevi\'c, J. 2011, ApJ, 738, 68

\bibitem[Popovi\'c  et al. (2004)]{po04} Popovi\'c, L. \v C., Mediavilla, E., Bon, E., Ili\'c, D. 
2004, A\&A, 423, 909 

\bibitem[Shapovalova et al.  (2012)]{sh12}
Shapovalova, A. I., Popovi\'c, L. \v C., Burenkov, A. N. et al. 2012, ApJS, 202, 10

\bibitem[Shen \& Ho (2014)]{sh14} Shen, Y.,  Ho, L. C. 2014, Nature, 513, 210 

\bibitem[Shields et al. (2010)]{sh10} Shields, G. A., Ludwig, R. R., Salviander, S. 
2010, ApJ, 721, 835 


\bibitem[Sriram et al. (2022)]{sr22} Sriram, K.,  Nour, D., Choi, C. S. 2022, MNRAS, 510, 3222

\bibitem[Sulentic et al. (2000)]{su00} Sulentic, J. W., Zwitter, T., Marziani, P., Dultzin-Hacyan, D. 
2000, ApJ, 536, L5 

\bibitem[Veron-Cetty et al. (2004)]{VC2004} Veron-Cetty M.-P., Joly M., Veron P., 2004, A\&A, 417, 515

\bibitem[Wills et al. (1985)]{wi1985} Wills, B. J., Netzer, H., Wills, D. 1985, ApJ, 288, 94

\bibitem[Wu \& Liu (2004)]{Wu2004}  Wu, X.-B. \& Liu, F.-K. 2004, ApJ, 614, 91

\end{thebibliography}
\end{document}